\documentclass[aps,prb,twocolumn,amsmath,amssymb,superscriptaddress,floatfix]{revtex4-1}

\usepackage{graphicx}
\usepackage{multirow}
\usepackage{color}
\usepackage{bm}

\def\ket#1{|#1\rangle }

\def\n{\nonumber \\ }
\def\tensor{\otimes}

\newcommand{\blue}[1]{{\textcolor{blue}{#1}}}

\begin{document}

\title{
Topological charges of three-dimensional Dirac semimetals with rotation symmetry
}

\author{Bohm-Jung Yang}
\affiliation{RIKEN, Center for Emergent Matter Science, Wako, Saitama, 351-0198, Japan}
\affiliation{Department of Physics and Astronomy, Seoul National University, Seoul 151-747, Korea}
\affiliation{Center for Correlated Electron Systems, Institute for Basic Science (IBS), Seoul 151-747, Korea}

\author{Takahiro Morimoto}
\affiliation{RIKEN, Center for Emergent Matter Science, Wako, Saitama, 351-0198, Japan}
\affiliation{Condensed Matter Theory Laboratory, RIKEN, Wako, Saitama, 351-0198, Japan}

\author{Akira Furusaki}
\affiliation{RIKEN, Center for Emergent Matter Science, Wako, Saitama, 351-0198, Japan}
\affiliation{Condensed Matter Theory Laboratory, RIKEN, Wako, Saitama, 351-0198, Japan}

\date{\today}
\begin{abstract}
In general, the stability of a band crossing point indicates the presence of
a quantized topological number associated with it.
In particular, the recent discovery of three-dimensional Dirac semimetals in Na$_{3}$Bi and Cd$_{3}$As$_{2}$
demonstrates that a Dirac point with four-fold degeneracy can be stable as long as certain
crystalline symmetries are supplemented in addition to the time-reversal and inversion symmetries. However,
the topological charges associated with Na$_{3}$Bi and Cd$_{3}$As$_{2}$ are not clarified yet.
In this work, we identify the topological charge of three-dimensional Dirac points.
It is found that although
the simultaneous presence of the time-reversal and inversion symmetries
forces the net chiral charge to vanish, a Dirac point can carry another quantized topological
charge when an additional rotation symmetry is considered.
Two different classes of Dirac semimetals are identified depending on the nature of the rotation symmetries.
First, the conventional symmorphic rotational symmetry which commutes with the inversion
gives rise to the class I Dirac semimetals having a pair of Dirac points on the rotation axes.
Since the topological charges of each pair of Dirac points have the opposite sign,
a pair-creation or a pair-annihilation is required to change the number of Dirac points in the momentum space.
On the other hand, the class II Dirac semimetals possess a single isolated Dirac point
at a time-reversal invariant momentum, which is protected by a screw rotation.
The non-symmorphic nature of screw rotations allows the anti-commutation relation between the
rotation and inversion symmetries, which enables to circumvent the doubling
of the number of Dirac points and create a single Dirac point at the Brillouin zone
boundary.
\end{abstract}

\pacs{75.10.Pq,75.10.Jm,64.70.Tg}
\maketitle
\section{\label{sec:intro} Introduction}
After the discovery of graphene, a class of materials, dubbed Dirac semimetals,
have come to the fore of condensed matter research.
In general, a Dirac semimetal has several Fermi points
around which pseudo-relativistic linear dispersion relation is realized.
This pseudo-relativistic energy dispersion forces the density of states
on the Fermi level to vanish without opening of an energy gap,
which is the unique property of Dirac semimetals distinct from ordinary metals or insulators~\cite{Graphene}.
In particular, the recent theoretical prediction~\cite{Na3Bi_DFT,Cd3As2_DFT}
and the experimental confirmation~\cite{Na3Bi_Exp1,Na3Bi_Exp2,Na3Bi_Exp3,Cd3As2_Exp1,Cd3As2_Exp2,Cd3As2_Exp3,Cd3As2_Exp4}
of three-dimensional (3D) Dirac semimetals in
Na$_{3}$Bi and Cd$_{3}$As$_{2}$
demonstrate that there are a variety of materials realizing
Dirac semimetals in both two dimensions and three dimensions.
Such a diversity of Dirac materials requires us to find
a systematic way to characterize and classify them.

In general, the stability of nodal points in a Dirac semimetal
has topological origin. This is because there is no characteristic
energy scale, such as the Fermi energy or the energy gap,
characterizing the perturbative stability of the system.
For instance, a nodal point in graphene
carries a quantized pseudo-spin winding number
that is defined on a loop encircling the Dirac point~\cite{Graphene}.
On the other hand, a nodal point in 3D Weyl semimetals
is endowed with a Chern number defined on a two-dimensional (2D)
closed surface surrounding the Weyl point~\cite{Wan}.
The presence of such a quantized topological charge carried by a nodal point
guarantees its stability, hence a nodal point can be annihilated
only by colliding with another nodal point with the opposite
topological charge as long as the symmetry of the system is preserved.
Recently, there have been several theoretical studies
which attempt to classify the topological invariants of nodal points,
and to extend the concept of topological band theory
to gapless systems such as semimetals and nodal
superconductors~\cite{Horava,Beri,nodalSC1,nodalSC2,nodalSC3,Manes,Matsuura,Zhao1,Zhao2,Chiu,BJYang,Morimoto_Z2,FanZhang,Young1,Young2,Sigrist,nonsymmorphicTI}.
However,
in our opinion, a proper definition of
the topological charge of Dirac points in Na$_{3}$Bi or Cd$_{3}$As$_{2}$
has not been given so far.

Dirac points in Na$_{3}$Bi or Cd$_{3}$As$_{2}$ are protected
by the time-reversal ($T$), the inversion ($P$),
and the rotation symmetries~\cite{Na3Bi_DFT,Cd3As2_DFT,BJYang}.
The experimental observation of Dirac points in these systems
demonstrates their stability, hence the presence of topological invariants associated with them.
Moreover, the theoretical observation of pair-annihilation
and pair-creation of Dirac points~\cite{BJYang,TPT}
indicates that the topological charges of the two Dirac points should have the opposite sign.
Then the question is what the nature of the topological charge is associated with the Dirac points.
Considering the two-dimensionality of the sphere surrounding a Dirac point,
the natural candidate is
either a Chern number similar to the case of Weyl semimetals,
or a $Z_{2}$ invariant associated with $T$ symmetry satisfying $T^{2}=-1$.
However, the simultaneous presence of the time-reversal and the inversion
symmetries forces the Berry curvature to be zero
at each momentum, hence the Chern number of the Dirac point, which
is basically the integral of the Berry curvature, also vanishes.
Moreover, a Dirac point can carry a $Z_{2}$ topological charge
only in the presence of SU(2) spin rotation symmetry
together with time-reversal and inversion symmetries
satisfying $(TP)^{2}=1$
as shown in Ref.~\onlinecite{Morimoto_Z2}.
These indicate that a special care is required
to find the topological charge of a Dirac point in Na$_{3}$Bi or Cd$_{3}$As$_{2}$,
which should obviously be distinct from
the monopole charge of Weyl semimetals.

In this work, we will show that the Dirac points in Na$_{3}$Bi or
Cd$_{3}$As$_{2}$ are characterized by topological invariants
of zero-dimensional subsystems defined on the rotation axis.
Since the rotation eigenvalue is a good quantum number on the rotation axis,
a zero-dimensional topological invariant can be defined
by comparing the rotation eigenvalues of the valence and conduction bands
at two points enclosing a Dirac point.
We find that the nature of 3D Dirac semimetals
strongly depends on the nature of the rotation symmetry.
Namely, the ordinary symmorphic rotation symmetry commuting with
the inversion symmetry always creates a pair of Dirac points
having the opposite topological charges, and generates class I Dirac semimetals.
Both Na$_{3}$Bi and Cd$_{3}$As$_{2}$ belong to this class.

On the other hand, we find a different class of Dirac semimetals
when the system has a screw rotation symmetry.
In general, the presence
of non-symmorphic symmetries, such as screw rotations and glide mirror symmetries,
guarantees a nontrivial band connection at the Brillouin zone
boundary~\cite{Zak1, Zak2, Zak3, Parameswaran}.
Also, it is proposed that when the double space group
of non-symmorphic crystals satisfies certain conditions,
a Dirac point can be realized at the Brillouin zone boundary~\cite{Young1}.
Consistent with these results, our theoretical study shows that when the band degeneracy
at the zone boundary is compatible with the time-reversal
and inversion symmetries, a single
isolated Dirac point can be created on the rotation axis.
Based on this observation, we define a class II Dirac semimetal
which is protected by a screw rotation symmetry and the inversion, which are mutually anti-commuting,
in addition to the time-reversal symmetry.
The partial translation associated with a screw rotation
adds a U(1) phase to the rotation eigenvalue,
which varies on the rotation axis.
This projective nature of a screw rotation enables to circumvent
the doubling of Dirac points, and create a single isolated Dirac point
at a time-reversal invariant momentum at the Brillouin zone boundary.

The rest of the paper is organized in the following way.
We describe the general idea to define a topological charge
in systems with rotation symmetry in Sec.~\ref{sec:general}.
Based on this general idea, class I Dirac semimetals
are defined and systematically classified
in Sec.~\ref{sec:class1}.
In particular, we show that the doubling of Dirac points
is unavoidable in class I Dirac semimetals.
Sec.~\ref{sec:class2_screwrotation} is about
the nontrivial band connection generated by non-symmorphic screw rotation symmetries.
Here we show that the partial translation associated with
a screw rotation induces a momentum dependent U(1) phase factor to the rotation eigenvalue,
which enables to circumvent the fermion number doubling and
protects a single Dirac point at the Brillouin zone boundary.
Based on the discussion in Sec.~\ref{sec:class2_screwrotation},
class II Dirac semimetals are defined and systematically classified
in Sec.~\ref{sec:class2}.
We present the conclusion and discussion in Sec.~\ref{sec:discussion}.
In the Appendix, we prove that there is no stable Dirac semimetal
in systems only with the time-reversal and inversion symmetries based
on $K$ theory approach.
The classification of Dirac semimetals in $C_{2}$ invariant systems shown
in the main text
is also confirmed by using the $K$ theory approach.
Finally, we present a short discussion about the stability of
2D Dirac semimetals protected by two-fold screw rotations.

\section{\label{sec:general} General idea: role of rotational symmetry in symmorphic crystals}
In general, electronic systems having only the time-reversal ($T$) and
inversion ($P$) symmetries cannot support a stable Dirac point
with a quantized topological charge~\cite{Murakami,Wan,Burkov}.
In Appendix A, we have revisited this known fact in a different perspective
and proved it by using $K$ theory approach.
Thus additional crystalline symmetries play a crucial role to stabilize Dirac semimetals
realized in Na$_{3}$Bi and Cd$_{3}$As$_{2}$.
Here we consider the role of the additional rotation symmetry ($C_{N}$)
in addition to $P$ and $T$.
For convenience, we first focus on 3D crystals with a symmorphic space group symmetry
in which the point group can be completely separable from pure translation operations.
Also we choose the $z$ axis
as the axis for $C_{N}$ rotation with $N$ indicating the discrete rotation angle of $2\pi/N$ ($N=2, 3, 4, 6$).
Under the operation of the $C_{N}$ symmetry, the Hamiltonian satisfies
\begin{align}
C_N H(k_x,k_y,k_z) (C_N)^{-1}&=H(\tilde k_x, \tilde k_y,k_z),
\end{align}
where $(\tilde k_x, \tilde k_y)$ is obtained from $2\pi/N$ rotation of $(k_x,k_y)$, i.e.,
$(\tilde{k}_{x}+i\tilde{k}_{y})=(k_{x}+ik_{y})e^{2\pi i/N}$.
The symmetry operators satisfy
\begin{align}
&T^2=-1,\quad  (C_N)^N=-1,\quad  P^{2}=1,
\end{align}
and
\begin{align}
&[T,P]=0,\quad [T,C_{N}]=0,\quad [P,C_{N}]=0,
\end{align}
where we have considered the fact that an electron is a spin-1/2 particle.

\begin{figure}[t]
\centering
\includegraphics[width=4.0 cm]{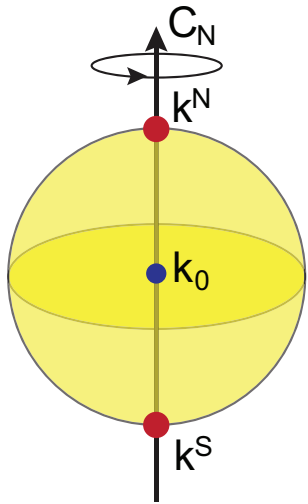}
\caption{
Local geometry around a Dirac point at $\bm{k}=\bm{k}_{0}$ sitting on the rotation axis.
The sphere surrounds a Dirac point at the center.
$\bm{k}^{N}$ and $\bm{k}^{S}$ mark the points on the sphere
crossing the rotation axis.
}
\label{fig:sphere}
\end{figure}
Now let us explain the general idea of how to determine the topological charge of a Dirac point locating
at a generic point $\bm{k}_0=(0,0,k_z^0)$ on the rotation axis.
To determine the topological charge,
we first consider a sphere in the momentum space surrounding the Dirac point at $\bm{k}=\bm{k}_0$
in a $C_N$ symmetric way as shown in Fig.~\ref{fig:sphere}. Namely,
the center of the sphere sits on the rotation axis.
At every point on the sphere, the Hamiltonian is invariant
under the compound antiunitary symmetry $PT$ satisfying $(PT)^{2}=-1$.
Moreover, the intersection of the $k_z$ axis and the sphere consists of
a north pole at $\bm{k}^N=(0,0,k_z^N)$ and a south pole at $\bm{k}^S=(0,0,k_z^S)$,
which are invariant under the rotation.
We define the topological charge of the Dirac point from the topological numbers
associated with these two points.
Since the Hamiltonian commutes with the rotation operator $C_{N}$ at these points,
\begin{align}
[H(\bm{k}^{N/S}), C_N]=0,
\end{align}
hence $H(\bm{k}^{N/S})$ can be block-diagonalized in the eigenspace of $C_{N}$
with the eigenvalues $J_{m}$ given by
\begin{align}\label{eqn:Jm}
J_m &= \exp \left( i\frac{2m+1}{N}\pi \right), \qquad m=0,\ldots,N-1.
\end{align}
Since the $PT$ symmetry is not satisfied in each $C_{N}$ eigenspace with a given $J_{m}$ in general,
(one exceptional case is shown in Sec.~\ref{sec:class1_odd}),
each diagonal block of $H(\bm{k}^{N,S})$ belongs to the symmetry class A in terms of
the Altland-Zirnbauer classification scheme~\cite{AZ},
hence carries an integer topological number $n^{N,S}_m$.
Here $n^N_m$ or $n^S_m$ indicates the topological invariant
of a zero-dimensional system belonging to the symmetry class A,
which is defined as
\begin{subequations}
\begin{align}\label{eqn:TopologicalCharge}
n^N_m &\equiv
\frac{1}{2}\!\left[
N_{\textrm{c}}(J_{m},\bm{k}^{N})-N_{\textrm{v}}(J_{m},\bm{k}^{N})
\right],
\\
n^S_m &\equiv
\frac{1}{2}\!\left[
N_{\textrm{c}}(J_{m},\bm{k}^{S})-N_{\textrm{v}}(J_{m},\bm{k}^{S})
\right],
\end{align}
\end{subequations}
where $N_{\textrm{c}/\textrm{v}}(J_{m},\bm{k})$ denotes the number of conduction bands ($\textrm{c}$) or valence bands ($\textrm{v}$)
with the eigenvalue $J_{m}$ at the momentum $\bm{k}$.
It is worth to note that a trivial conduction or valence band with a constant energy can always be added
to each $J_m$ sector,
so that the sum $n^N_m+n^S_m$ can be changed freely.
Therefore the nontrivial topological number in the $J_m$ sector is determined by the difference
\begin{align}
\nu_m\equiv n^N_m-n^S_m.
\end{align}

Next let us consider possible constraints
to the allowed $\nu_{m}$ values.
For a gapped system defined on the sphere, the number of
conduction bands and that of valence bands are constants independent of
the momentum on the sphere, which leads to the following constraint
\begin{align}\label{eqn:constraint_total}
\sum_m \nu_m&=\sum_m \big(n^N_m-n^S_m\big)
\nonumber\\
&=\sum_m\big\{[N_{\textrm{c}}(J_{m},\bm{k}^{N})
-N_{\textrm{c}}(J_{m},\bm{k}^{S})]
\nonumber\\
&\qquad\quad
-[N_{\textrm{v}}(J_{m},\bm{k}^{N})-N_{\textrm{v}}(J_{m},\bm{k}^{S})]\big\}
\nonumber\\
&=0.
\end{align}
Moreover, since $PT$ symmetry imposes additional constraints
between different $\nu_{m}$ values,
the number of independent topological invariants depends
on the details of the symmetry
as shown in Sec.~\ref{sec:class1} and \ref{sec:class2}.
However, as long as a Dirac point possesses a nonzero $\nu_{m}$ value,
it guarantees the stability of the relevant Dirac point.
Hence the set of nonzero $\nu_{m}$ can be considered as a topological invariant
characterizing a stable Dirac point.

\section{\label{sec:class1} Class I Dirac semimetals}

Class I Dirac semimetals are protected by
the ordinary rotation symmetry commuting with
an inversion symmetry, i.e.,
\begin{eqnarray}
[P,C_{N}]=0.
\end{eqnarray}
Let us consider the eigenstate $\ket{\psi_m}$ of the $C_{N}$ operator
with the eigenvalue $J_{m}$.
The $PT$ symmetry requires
\begin{align}
C_N PT \ket{\psi_m} &= PT C_N \ket{\psi_m} \n
&= PT J_m \ket{\psi_m} \n
&= J_m^* PT \ket{\psi_m} \n
&= J_{N-m-1} PT \ket{\psi_m},
\end{align}
from which we find $PT\ket{\psi_m}$ is an eigenstate of $C_{N}$
with the eigenvalue $J_{N-m-1}$.
Therefore when a state with the eigenvalue $J_{m}$ is occupied (or unoccupied),
there should be another occupied (or unoccupied) state
with the eigenvalue $J_{N-m-1}$,
which leads to the constraint
\begin{align}\label{eqn:constraint_1}
\nu_{m}=\nu_{N-m-1}.
\end{align}
For further analysis, we distinguish two cases
based on the parity of $N$ as shown below.
\begin{figure}[t]
\centering
\includegraphics[width=8.5 cm]{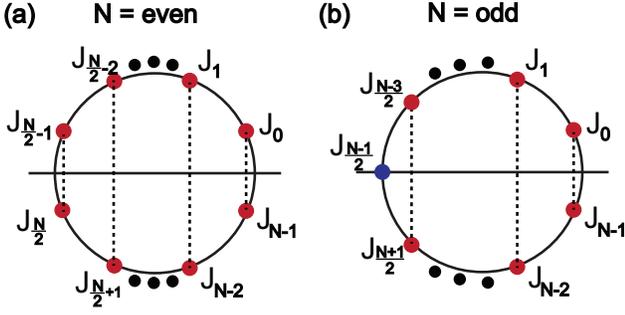}
\caption{
Constraints on the rotation eigenvalues (a) for even $N$, (b) for odd $N$.
The (black) solid circle indicates a unit circle in the complex plane,
and each (red) dot on the circle denotes $J_{m}$.
Two dots connected by a dotted line are related by the $PT$ symmetry,
hence only one of them is independent.
}
\label{fig:commuting}
\end{figure}

\subsection{\label{sec:class1_even} $C_{N}$ symmetric systems with even $N$}
Due to the $PT$ symmetry,
the $C_{N}$ eigenspaces with the eigenvalues $J_{m}$
and $J_{N-m-1}$ can be paired as $\{J_m, J_{N-m-1}\}$
with $m=0,\ldots, N/2-1$.
[See Fig.~\ref{fig:commuting} (a).]
Since $PT$ interchanges eigenspaces within each pair,
$PT$ is not a symmetry in each eigenspace separately, and
each eigenspace belongs to the symmetry class A.
Therefore an integer topological invariant defined
in Eq.~(\ref{eqn:TopologicalCharge}) can be computed
in each eigenspace with $J_{m}$.
Considering the constraints shown in Eq.~(\ref{eqn:constraint_total}) and Eq.~(\ref{eqn:constraint_1}),
we conclude that the topological charge of the Dirac point is given by
\begin{align}
(\nu_0,\ldots, \nu_{\frac{N}{2}-2}) \in \mathbb{Z}^{\frac{N}{2}-1}.
\end{align}
Therefore, the topological charge of a Dirac point with $C_{4}$ or $C_{6}$
symmetry is an element of $\mathbb{Z}$ or $\mathbb{Z}^2$, respectively.
At the same time, it implies that
a $C_{2}$ invariant system cannot support a stable Dirac point.

\subsection{\label{sec:class1_odd} $C_{N}$ symmetric systems with odd $N$ }
When $N$ is odd, the $PT$ symmetry pairs the eigenspaces of $C_{N}$
in a slightly different way as compared to even $N$ cases
as shown in Fig.~\ref{fig:commuting} (b).
At first, we find $\frac{1}{2}(N-1)$ pairs of eigenspaces
\begin{align}
\{J_m,J_{N-m-1}\}, \qquad m=0,\ldots, \frac{N-3}{2},
\end{align}
On the other hand, the remaining eigenspace with the eigenvalue
$J_{(N-1)/2}$ is invariant under the $PT$ symmetry,
hence belongs to class AII.
Thus, in a block-diagonalized Hamiltonian $H(\bm{k}^{N/S})$,
there are $\frac{1}{2}(N-1)$ blocks belonging to class A
and an extra block with the eigenvalue $J_{(N-1)/2}$ belonging
to class AII.
In both symmetry classes, zero dimensional systems have
an integer topological invariant, which is defined as
the difference in the number of conduction bands and that of valence bands.
Hence as in the even $N$ case, the topological charge can be defined as
\begin{align}
\nu_m=n^N_m-n^S_m,
\end{align}
in each eigenspace with the eigenvalue $J_m$.
The constraints to the topological numbers $\nu_{m}$ are
\begin{align}
(i)~& \nu_m = \nu_{N-m-1}, \quad m=0,\ldots, \frac{N-3}{2}, \\
(ii)~& \sum_m \nu_m =0.
\end{align}
Thus the independent topological charge for a Dirac point is given by
\begin{align}
(\nu_0,\ldots, \nu_{(N-3)/2}) \in \mathbb{Z}^{\frac{N-1}{2}}.
\end{align}
Therefore a Dirac point with $C_{3}$ symmetry
has an integer ($\mathbb{Z}$) topological charge.
\begin{table}[h]
\begin{tabular}{c c c}
\hline
\hline
$C_{N}$ & & Topological charge \\
\hline
\hline
$C_{2}$ & & Not allowed\\
$C_{3}$ & & $\mathbb{Z}$ \\
$C_{4}$ & & $\mathbb{Z}$ \\
$C_{6}$ & & $\mathbb{Z}\times\mathbb{Z}$  \\
\hline \hline
\end{tabular}
\caption{Summary of topological charges of class I Dirac semimetals.}
\end{table}\label{table:1}

\subsection{Applications: classification of stable Dirac points in 4-band systems}
Let us apply the general theory developed above to minimal 4-band models,
and classify stable Dirac points.
In a 4-band model, a pair of doubly degenerate bands cross
at a Dirac point which we assume to sit on the Fermi level.
On the rotation axis $\bm{k}=(0,0,k_{z})$,
each band is assigned with a quantum number $J_{m}$.
Since the pair of degenerate bands should have different rotation eigenvalues
to generate a stable Dirac point, each band with the rotation eigenvalue
$J_{m}$ satisfies $n^{N}_{m}=-n^{S}_{m}$.
Namely, a band which is below (above) the Fermi level at the momentum
$\bm{k}^{N}$ should be above (below) the Fermi level
at the momentum $\bm{k}^{S}$ to have a Dirac point in between.

\subsubsection{$C_{2}$ symmetric systems}

$J_{m=0}=\exp(i\pi/2)=i$ and $J_{m=1}=\exp(3i\pi/2)=-i=J_{m=-1}$
are the only allowed $C_{2}$ eigenvalues. Due to the $PT$ symmetry,
a pair of eigenstates
$\{|\psi_{m=0}(\bm{k})\rangle, |\psi_{m=1}(\bm{k})\rangle\}$
are always degenerate locally at each momentum $\bm{k}$,
hence $n_{m=0}^{N,S}=n_{m=1}^{N,S}$ and $N_{m=0}=N_{m=1}$.
Then a 4-band model can be constructed by introducing
two pairs of eigenstates
$\{|\psi_{m=0}^{A}(\bm{k})\rangle, |\psi_{m=1}^{A}(\bm{k})\rangle\}$
and $\{|\psi_{m=0}^{B}(\bm{k})\rangle, |\psi_{m=1}^{B}(\bm{k})\rangle\}$,
where $A, B$ indicates the valence band ($\textrm{v}$) or
the conduction band ($\textrm{c}$), respectively.
It is straightforward to show that $n_{m}^{N,S}=0$ ($m=$ 0, 1),
because if one state is occupied,
among $\{|\psi_{m}^{A}(\bm{k})\rangle$,
$|\psi_{m}^{B}(\bm{k})\rangle\}$, the other state is unoccupied.
Therefore $\nu_{m=0, 1}=0$ and there is no stable Dirac point
with a nontrivial topological invariant in systems with
$C_{2}$ symmetry.

\subsubsection{$C_{3}$ symmetric systems}

Possible $C_{3}$ eigenvalues are
$J_{m=0}=\exp(i\pi/3)$, $J_{m=1}=\exp(i\pi)$,
$J_{m=2}=\exp(5i\pi/3)$.
Due to the $PT$ symmetry,
$\{|\psi_{m=0}(\bm{k})\rangle, |\psi_{m=2}(\bm{k})\rangle\}$ and
$\{|\psi_{m=1}(\bm{k})\rangle, |\tilde{\psi}_{m=1}(\bm{k})\rangle\}$
form degenerate pairs,
where $|\tilde{\psi}_{m=1}(\bm{k})\rangle=PT|\psi_{m=1}(\bm{k})\rangle$.
Thus $\nu_{m=0}=\nu_{m=2}$.
Since $\nu_{m=0}+\nu_{m=1}+\nu_{m=2}=0$, we have
\begin{eqnarray}
\overline{\nu}_{1}\equiv \nu_{m=0}=\nu_{m=2}=-\frac{1}{2}\nu_{m=1}.
\end{eqnarray}
Hence there is only one independent topological number,
$\overline{\nu}_{1}\in \mathbb{Z}$.
Since the topological charge of a Dirac point can be nonzero
only when the valence and conduction bands have different rotation eigenvalues,
a 4-band model can be constructed by using a basis
$\{|\psi_{m=0}^{A}(\bm{k})\rangle, |\psi_{m=2}^{A}(\bm{k})\rangle,
|\psi_{m=1}^{B}(\bm{k})\rangle, |\tilde{\psi}_{m=1}^{B}(\bm{k})\rangle\}$
where $A=\textrm{v}$ $(\textrm{c})$ and $B=\textrm{c}$ $(\textrm{v})$.
Since $n_{m=0}^{N,S}=n_{m=2}^{N,S}=-\frac{1}{2}n_{m=1}^{N,S}=\pm\frac12$
and $n_{m}^{N}=-n_{m}^{S}$ for 4-band models,
the Dirac point has a nonzero topological invariant
\begin{eqnarray}
\overline{\nu}_{1}=\pm1.
\end{eqnarray}

\subsubsection{$C_{4}$ symmetric systems}

Possible $C_{4}$ eigenvalues are
$J_{m=0}=\exp(i\pi/4)$, $J_{m=1}=\exp(3i\pi/4)$,
$J_{m=2}=\exp(5i\pi/4)$, $J_{m=3}=\exp(7i\pi/4)$.
Due to the $PT$ symmetry,
$\{|\psi_{m=0}(\bm{k})\rangle, |\psi_{m=3}(\bm{k})\rangle\}$
and $\{|\psi_{m=1}(\bm{k})\rangle, |\psi_{m=2}(\bm{k})\rangle\}$
form degenerate pairs
at each momentum, thus $\nu_{m=0}=\nu_{m=3}$ and $\nu_{m=1}=\nu_{m=2}$.
Since $\sum_{m}\nu_{m}=0$,
\begin{eqnarray}
\overline{\nu}_{1}\equiv \nu_{m=0}=\nu_{m=3}=-\nu_{m=1}=-\nu_{m=2},
\end{eqnarray}
hence there is only one independent topological number, $\overline{\nu}_{1}$.
Since the topological charge of a Dirac point can be nonzero
only when the valence and conduction bands have different rotation eigenvalues,
a 4-band model with Dirac points can be constructed by using a basis
$\{|\psi_{m=0}^{A}(\bm{k})\rangle, |\psi_{m=3}^{A}(\bm{k})\rangle,
|\psi_{m=1}^{B}(\bm{k})\rangle, |\psi_{m=2}^{B}(\bm{k})\rangle\}$
where $A=\textrm{v}$ $(\textrm{c})$ and $B=\textrm{c}$ $(\textrm{v})$.
Since
$n_{m=0}^{N,S}=n_{m=3}^{N,S}=-n_{m=1}^{N,S}=-n_{m=2}^{N,S}=\pm\frac12$
and $n_{m}^{N}=-n_{m}^{S}$ for 4-band models,
the Dirac point has a nonzero topological invariant
\begin{eqnarray}
\overline{\nu}_{1}=\pm1.
\end{eqnarray}

\subsubsection{$C_{6}$ symmetric systems}

In the presence of a $C_6$ rotation symmetry,
$\{|\psi_{m=0}(\bm{k})\rangle, |\psi_{m=5}(\bm{k})\rangle\}$,
$\{|\psi_{m=1}(\bm{k})\rangle, |\psi_{m=4}(\bm{k})\rangle\}$, and
$\{|\psi_{m=2}(\bm{k})\rangle, |\psi_{m=3}(\bm{k})\rangle\}$
form degenerate pairs.
Thus $\nu_{m=0}=\nu_{m=5}$, $\nu_{m=1}=\nu_{m=4}$, $\nu_{m=2}=\nu_{m=3}$.
Considering $\sum_{m}\nu_{m}=0$, we can find two independent topological numbers $(\overline{\nu}_{1},\overline{\nu}_{2})\in \mathbb{Z}^{2}$,
which, for instance, can be defined as,
\begin{eqnarray}
&&\overline{\nu}_{1}\equiv \nu_{m=0}=\nu_{m=5},
\nonumber\\
&&\overline{\nu}_{2}\equiv \nu_{m=1}=\nu_{m=4}.
\end{eqnarray}
However, for convenience, we can also use $(\nu_{m=0}, \nu_{m=1}, \nu_{m=2})$ to
indicate the topological charge in which $\nu_{m=0}+\nu_{m=1}+\nu_{m=2}=0$.
A 4-band model can be constructed by choosing two different pairs of eigenstates
such as
\begin{eqnarray}\label{eqn:4band_C6}
&&\{|\psi_{m=0}\rangle, |\psi_{m=5}\rangle\, |\psi_{m=1}\rangle, |\psi_{m=4}\rangle\},
\nonumber\\
&&\{|\psi_{m=0}\rangle, |\psi_{m=5}\rangle\, |\psi_{m=2}\rangle, |\psi_{m=3}\rangle\},
\\
&&\{|\psi_{m=1}\rangle, |\psi_{m=4}\rangle\, |\psi_{m=2}\rangle, |\psi_{m=3}\rangle\}.
\nonumber
\end{eqnarray}
For a given 4-band model, a nonzero topological number $\nu_{m}=\pm 1$ can be assigned if $J_{m}$
is the eigenvalue of one of the four bands. Whereas $\nu_{m}=0$ if $J_{m}$ is the eigenvalue
of the other two states which are not included in the 4-band model.
Therefore the topological charges of the system are in the form of
\begin{eqnarray}
&&(\nu_{m=0}, \nu_{m=1}, \nu_{m=2})=(\pm1,\mp1,0),
\nonumber\\
&&(\nu_{m=0}, \nu_{m=1}, \nu_{m=2})=(\pm1,0,\mp1),
\\
&&(\nu_{m=0}, \nu_{m=1}, \nu_{m=2})=(0,\pm1,\mp1),
\nonumber
\end{eqnarray}
for each case shown in Eq.~(\ref{eqn:4band_C6}), respectively.
Then the corresponding $(\overline{\nu}_{1},\overline{\nu}_{2})$ are
\begin{eqnarray}
&&(\overline{\nu}_{1}, \overline{\nu}_{2})=(\pm1,\mp1),
\nonumber\\
&&(\overline{\nu}_{1}, \overline{\nu}_{2})=(\pm1,0),
\\
&&(\overline{\nu}_{1}, \overline{\nu}_{2})=(0,\pm1),
\nonumber
\end{eqnarray}
respectively.

\begin{table}[h]
\begin{tabular}{c c c c c}
\hline
\hline
$C_{N}$  & & 4-band model & & Materials \\
\hline
\hline
$C_{2}$  & & Not allowed & &\\
$C_{3}$ & & $\overline{\nu}_{1}=\pm1$ & &Na$_{3}$Bi~[\onlinecite{Na3Bi_DFT}], strained TlN~[\onlinecite{DFT_TlN}]\\
$C_{4}$ & & $\overline{\nu}_{1}=\pm1$ & &Cd$_{3}$As$_{2}$~[\onlinecite{Cd3As2_DFT}]\\
$C_{6}$ & & $(\overline{\nu}_{1},\overline{\nu}_{2})=(\pm1,\mp1)$& & \\
$C_{6}$ & & $(\overline{\nu}_{1},\overline{\nu}_{2})=(\pm1,0)$& & \\
$C_{6}$ & & $(\overline{\nu}_{1},\overline{\nu}_{2})=(0,\pm1)$& & \\
\hline \hline
\end{tabular}
\caption{Topological charges of class I Dirac semimetals
for 4-band models and relevant materials.
}
\end{table}\label{table:2}

\subsection{\label{subsec:doubling1} Fermion number doubling in class I Dirac semimetals}
Up to now, we have described how to determine the topological charge of
a single Dirac point. Now let us compare the topological charges of two Dirac points at the momenta $\bm{k}_{0}$ and $-\bm{k}_{0}$.
Due to the inversion symmetry satisfying $[P,C_{N}]=0$, the eigenstates at $\bm{k}$ and $-\bm{k}$ satisfy
the following relationship,
\begin{align}
C_{N} P \ket{\psi_{m}(\bm{k})} &= P C_{N} \ket{\psi_{m}(\bm{k})} \n
&= P J_{m} \ket{\psi_{m}(\bm{k})} \n
&= J_{m} P \ket{\psi_{m}(\bm{k})},
\end{align}
which means that if there is an eigenstate with the eigenvalue $J_{m}$ at $\bm{k}$,
there should be a degenerate eigenstate with the same eigenvalue $J_{m}$ at $-\bm{k}$.
This imposes the following condition of
\begin{align}
n^{N,S}_{m}(\bm{k}_{0})=n^{S,N}_{m}(-\bm{k}_{0}).
\end{align}
It is to be noted that the north (south) pole at $\bm{k}_{0}$ and
the south (north) pole at $-\bm{k}_{0}$
are interchanged
under the inversion symmetry.
Thus we obtain
\begin{align}
\nu_{m}(\bm{k}_{0})=-\nu_{m}(-\bm{k}_{0}).
\end{align}
Since the net topological charge of the two Dirac points related by the inversion symmetry
is zero, we obtain the following conclusions.
\begin{itemize}
\item The total topological charge of two Dirac points within the first Brillouin zone
should be zero in each angular momentum channel ($J_{m}$), i.e.,
\begin{align}\label{item:NM}
\sum_{i_{D}} \nu_{m}^{(i_{D})}=0,
\end{align}
where $i_{D}$ labels Dirac points.
It is worth to note that this is nothing but the Nielsen-Ninomiya theorem~\cite{NM1,NM2}
for three-dimensional Dirac semimetals.
\item A stable Dirac point with a nontrivial topological charge cannot exist
at a time-reversal invariant momentum (TRIM) where $\bm{k}_{0}=-\bm{k}_{0}$
modulo a reciprocal lattice vector due to the relationship
\begin{align}\label{item:TRIM}
\nu_{m}(\bm{k}_{0})=-\nu_{m}(-\bm{k}_{0})=-\nu_{m}(\bm{k}_{0})=0.
\end{align}
\end{itemize}
\section{\label{sec:class2_screwrotation} Screw rotations and a single Dirac point}

\subsection{Projective symmetry and circumventing fermion number doubling}
\begin{figure}[t]
\centering
\includegraphics[width=8 cm]{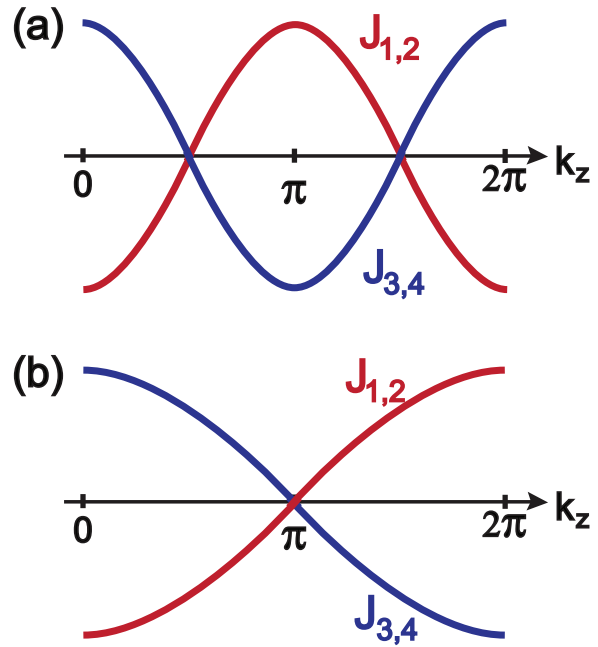}
\caption{
Band structure along the rotation axis ($z$ axis)
of (a) a class I Dirac semimetal and (b) a class II Dirac semimetal.
$J_{1,2}$ and $J_{3,4}$ are the rotation eigenvalues of each doubly degenerate band.
A band crossing requires $J_{1,2}\neq J_{3,4}$.
In class I (II) Dirac semimetals, the band crossing condition
and the periodicity of the eigenstates are compatible (incompatible)
when $J_{1,2,3,4}$ are constant
on the rotation axis.
}
\label{fig:bandstructure}
\end{figure}

It is worth to note that the doubling of the number of Dirac points
in class I Dirac semimetals results from
the commutation relation $[P,C_{N}]=0$ as discussed in Sec.~\ref{subsec:doubling1}.
This means that it may be possible to avoid the doubling of the Dirac points,
once the commutation relation is violated, i.e., $[P,C_{N}]\neq0$.
However, the presence of a single Dirac point
on the rotation axis brings about a more fundamental problem,
when the periodic structure of the system is considered.
This is because
the band crossing (or a nonzero topological charge of a Dirac point)
requires that the valence band and the conduction
band should have distinct eigenvalues, whereas the lattice periodicity requires
the continuity of the eigenstate and its relevant eigenvalues as described
in Fig.~\ref{fig:bandstructure}.
Therefore the presence of a single Dirac point or an odd number of Dirac points on the rotation axis sounds unphysical,
when the rotation symmetry exists along a line satisfying the periodic boundary condition.

One possible way to circumvent the contradiction is when the rotation symmetry is realized projectively.
Namely, if the rotation eigenvalue is well-defined only up to an additional phase factor,
it is possible to create a single Dirac point compatible with the lattice periodicity
by adjusting the phase degrees of freedom on the rotation axis.
In fact, a screw rotation is such an example of projective symmetry,
which can support a single isolated Dirac point as discussed in detail below.

\begin{figure}[t]
\centering
\includegraphics[width=8.5 cm]{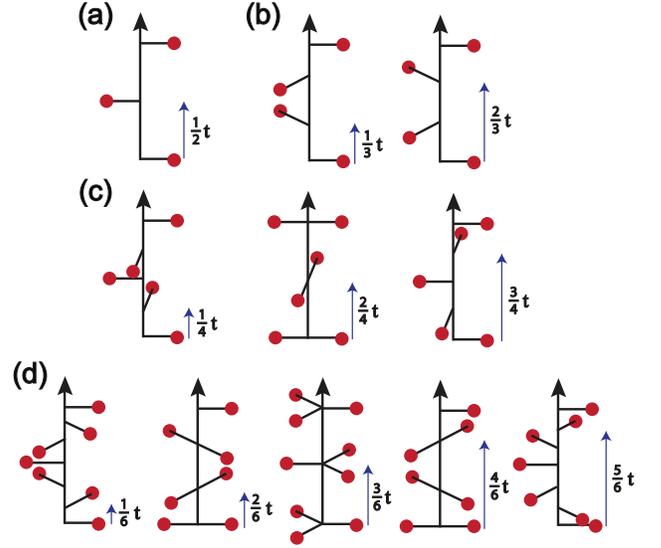}
\caption{
Schematic figure describing all possible screw rotations $\widetilde{C}_{N,p}$ in 3D crystals.
In each figure, the system is periodic under the lattice translation $\bm{t}$ along
the vertical direction. $\widetilde{C}_{N,p}$ indicates a $\frac{2\pi}{N}$ counterclockwise rotation
combined with a partial translation $\frac{p}{N}\bm{t}$.
(a) $\widetilde{C}_{2,1}$ symmetry.
(b) $\widetilde{C}_{3,p}$ symmetry ($p=1,2$).
(c) $\widetilde{C}_{4,p}$ symmetry ($p=1,2,3$).
(d) $\widetilde{C}_{6,p}$ symmetry ($p=1,2,3,4,5$).
}
\label{fig:screwrotations}
\end{figure}

A screw rotation ($\widetilde{C}_{N,p}$) is a non-symmorphic symmetry operation composed
of an ordinary rotation ($C_{N}$) followed by a partial lattice translation $\bm{\tau}_{p}=\frac{p}{N}\hat{z}$ ($p=1,...,N-1$)
parallel to the rotation axis.
Here $\hat{z}$ is the unit lattice translation along the $z$ axis assuming
that the screw axis is parallel to it.
Schematic figures describing all possible screw rotations in 3D crystals
are shown in Fig.~\ref{fig:screwrotations}.
Let us note that, in many crystals, the screw rotation axis does not pass the reference point of the point group symmetry, which is invariant under
point group operations of the lattice. In this case, the partial translation $\bm{\tau}_{p}$ associated with the screw rotation $\widetilde{C}_{N,p}$
also includes in-plane translation components perpendicular to the screw axis direction.
Generally, $\widetilde{C}_{N,p}$ can be compactly represented as
\begin{align}
\widetilde{C}_{N,p}=\{C_{N}|\bm{\tau}_{p}\},
\end{align}
where
\begin{align}
\bm{\tau}_{p}=\Big(\tau_{p,x},\tau_{p,y},\tau_{p,z}=\frac{p}{N}\Big).
\end{align}
In the real space, $\widetilde{C}_{N,p}$ transforms the spatial coordinates in the following way,
\begin{eqnarray}
\widetilde{C}_{N,p}: (x,y,z)\rightarrow (x'+\tau_{p,x},y'+\tau_{p,y},z+\tau_{p,z}),
\end{eqnarray}
where
\begin{eqnarray}
&&x'=x\cos\frac{2\pi}{N}-y\sin\frac{2\pi}{N},
\nonumber\\
&&y'=x\sin\frac{2\pi}{N}+y\cos\frac{2\pi}{N}.
\end{eqnarray}

Now we consider the combination of a screw rotation and the inversion symmetry.
At first, we see
\begin{eqnarray}\label{eqn:PRn1}
P\widetilde{C}_{N,p}: (x,y,z)\rightarrow (-x'-\tau_{p,x},-y'-\tau_{p,y},-z-\tau_{p,z}),
\nonumber
\end{eqnarray}
thus
\begin{eqnarray}\label{eqn:PRn2}
P\widetilde{C}_{N,p}=\{PC_{N}|-\bm{\tau}_{p}\}.
\end{eqnarray}
Similarly,
\begin{eqnarray}\label{eqn:RnP1}
\widetilde{C}_{N,p}P: (x,y,z)\rightarrow (-x'+\tau_{p,z},-y'+\tau_{p,z},-z+\tau_{p,z}),
\nonumber
\end{eqnarray}
thus
\begin{eqnarray}\label{eqn:RnP2}
\widetilde{C}_{N,p}P=\{PC_{N}|\bm{\tau}_{p}\}.
\end{eqnarray}
Let us note that $[P,C_{N}]=0$ in general.
Equations~(\ref{eqn:PRn2}) and (\ref{eqn:RnP2})
clearly show that generally $[P,\widetilde{C}_{N,p}]\neq0$ due to the partial translation
$\bm{\tau}_{p}$,
thus there is a chance to avoid the doubling of the Dirac points.

In the presence of the screw rotation symmetry $\widetilde{C}_{N,p}$,
the Bloch Hamiltonian $H(k_{z})$ on the rotation axis ($k_x=k_y=0$)
satisfies
\begin{align}
\widetilde{C}_{N,p}(k_{z})H(k_{z})\widetilde{C}_{N,p}^{-1}(k_{z})=H(k_{z}),
\end{align}
where
\begin{align}
[\widetilde{C}_{N,p}(k_{z})]^{N}=-\exp(-ipk_{z}).
\end{align}
Here the minus sign stems from the spin 1/2 nature of electrons.
Therefore all bands on the $k_{z}$ axis can be labeled
by the eigenvalues of $\widetilde{C}_{N,p}(k_{z})$ given by
\begin{align}\label{eqn:screweigenvalue}
\widetilde{J}_{m}(k_{z})&=\exp\Big[i\pi\frac{(2m+1)}{N}\Big]\exp(-i\frac{p}{N}k_{z})
\nonumber\\
&=J_{m}\exp\Big(-i\frac{p}{N}k_{z}\Big),
\end{align}
where $J_{m}$ is an eigenvalue of $C_{N}$ defined in Eq.~(\ref{eqn:Jm}).
It is worth to note that the eigenvalue of the screw rotation $\widetilde{C}_{N,p}$
is not $J_{m}$ but $J_{m}\exp(-i\frac{p}{N}k_{z})$
which varies along the rotation axis.
Therefore through the variation of this additional phase factor,
it may be possible to satisfy the condition for the band crossing to create a Dirac point and
the periodicity (or the continuity) of the eigenvalues, simultaneously,
even in the presence of a single Dirac point.
In fact, the assignment of non-quantized quantum numbers to fermions, such as $\widetilde{J}_{m}(k_{z})$
varying in the momentum space,
is one way to get around the fermion doubling problem,
as pointed out by Nielsen and Ninomiya in their seminal work~\cite{NM1,NM2}.

\subsection{Screw rotations and band connections at the zone boundary}

\begin{figure}[t]
\centering
\includegraphics[width=8.5 cm]{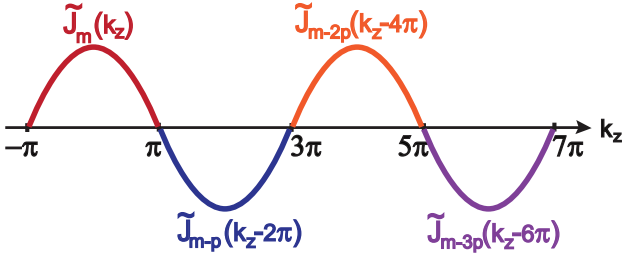}
\caption{
The band connection required by a screw rotation $\widetilde{C}_{N,p}$
when the system is $2\pi$ periodic along the $k_{z}$ direction.
}
\label{fig:bandconnection_general}
\end{figure}

The momentum dependence of screw rotation eigenvalues $\widetilde{J}_{m}(k_{z})$ shown in
Eq.~(\ref{eqn:screweigenvalue}) induces
nontrivial band connections between different eigenstates at the Brillouin zone boundary.
For instance, if the system has $2\pi$ periodicity along the $k_{z}$ axis,
we find that
\begin{align}\label{eqn:bandconnection}
\widetilde{J}_{m}(k_{z}+2\pi)&=\exp\Big[i\pi\frac{(2m-2p+1)}{N}\Big]\exp\Big(-i\frac{p}{N}k_{z}\Big)
\nonumber\\
&=\widetilde{J}_{m-p}(k_{z}),
\end{align}
thus the eigenstate with the eigenvalue $\widetilde{J}_{m}(k_{z})$
should be smoothly connected to the other eigenstate with the eigenvalue
$\widetilde{J}_{m-p}(k_{z})$ ($\widetilde{J}_{m+p}(k_{z})$) at the Brillouin zone boundary
$k_{z}=\pi$ ($k_{z}=-\pi$) as shown in Fig.~\ref{fig:bandconnection_general}.
This naturally gives rise to a band crossing point
at the Brillouin zone boundary.
If this band connection is compatible with the $T$ and $P$ symmetries,
a single Dirac point can be realized at the Brillouin zone boundary.
The $PT$ symmetry requires that
the state with the eigenvalue $\widetilde{J}_{m}(k_{z})=J_{m}\exp(-i\frac{p}{N}k_{z})$
should be locally degenerate with the other state
with the eigenvalue $\widetilde{J}_{N-m-1}(k_{z})=J^{*}_{m}\exp(-i\frac{p}{N}k_{z})$
at each $k_{z}$.
Similarly, we can expect the degeneracy between two states
with the eigenvalues $\widetilde{J}_{m-p}(k_{z})=J_{m-p}\exp(-i\frac{p}{N}k_{z})$
and $\widetilde{J}_{N-m+p-1}(k_{z})=J^{*}_{m-p}\exp(-i\frac{p}{N}k_{z})$, respectively.
Here the important point is that the screw rotation requires a nontrivial band connection
between $\widetilde{J}_{N-m-1}(k_{z})$ and
$\widetilde{J}_{N-m+p-1}(k_{z})$, similar to the relation shown in Eq.~(\ref{eqn:bandconnection}).
Namely,
\begin{align}
\widetilde{J}_{N-m-1}(k_{z}+2\pi)&=\widetilde{J}_{N-m-1-p}(k_{z})
\nonumber\\
&=\widetilde{J}_{N-m+p-1}(k_{z}),
\end{align}
which gives
\begin{align}
N-m-1-p=N-m+p-1 \quad(\text{mod}~N),
\end{align}
thus
\begin{align}\label{eqn:p_2pi}
p=\frac{N}{2}.
\end{align}
Since $p$ is an integer, this condition can be satisfied only in systems with
$\widetilde{C}_{2,1}$, $\widetilde{C}_{4,2}$, $\widetilde{C}_{6,3}$ symmetries.

Let us note that, in 3D crystals, the periodicity along the $k_{z}$ direction
can be longer than $2\pi/a_{z}$ although the system is periodic under
the translation by $a_{z}$ along the $z$ direction, unless $a_{z}$ is a primitive lattice vector.
(For instance, it happens in the face centered cubic lattice.)
Generally, when the system is $2n\pi$ periodic along the $k_{z}$ axis with an integer $1<n<N$,
\begin{align}\label{eqn:bandconnection_general}
\widetilde{J}_{m}(k_{z}+2n\pi)&=\exp\Big[i\pi\frac{(2m-2np+1)}{N}\Big]\exp\Big(-i\frac{p}{N}k_{z}\Big)
\nonumber\\
&=\widetilde{J}_{m-np}(k_{z}),
\end{align}
thus the eigenstate with the eigenvalue $\widetilde{J}_{m}(k_{z})$
should be smoothly connected to the other eigenstate with the eigenvalue
$\widetilde{J}_{m-np}(k_{z})$ ($\widetilde{J}_{m+np}(k_{z})$) at the Brillouin zone boundary
$k_{z}=n\pi$ ($k_{z}=-n\pi$).
Considering the $PT$ symmetry and following the same procedure
that we have used to derive Eq.~(\ref{eqn:p_2pi}),
we obtain
\begin{align}\label{eqn:p_2npi}
2np=0 \quad(\text{mod}~N).
\end{align}
For example, when the system is $4\pi$ periodic ($n=2$), Eq.~(\ref{eqn:p_2npi}) can also be satisfied
in systems with $\widetilde{C}_{4,1}$,
$\widetilde{C}_{4,2}$, $\widetilde{C}_{4,3}$, $\widetilde{C}_{6,3}$ symmetries.
However, in systems with $\widetilde{C}_{4,2}$ and $\widetilde{C}_{6,3}$ symmetries,
a $4\pi$ shift merely maps an eigenstate into itself, hence nontrivial
band connection at the Brillouin zone boundary is not expected.
On the other hand, when the system is $6\pi$ periodic ($n=3$),
Eq.~(\ref{eqn:p_2npi}) can be satisfied
in systems with $\widetilde{C}_{6,p}$ symmetry where $p=1,2,3,4,5$.
However, in the case of $\widetilde{C}_{6,2}$ and $\widetilde{C}_{6,4}$ symmetries, a $6\pi$ shift
connects an eigenstate with itself. Also in the $\widetilde{C}_{6,3}$ symmetric case,
a $6\pi$ shift is simply equivalent to  a $2\pi$ shift, which is already
considered before. Hence only the systems with $\widetilde{C}_{6,1}$
and $\widetilde{C}_{6,5}$ can support a nontrivial
band connection at the Brillouin zone boundary $k_{z}=\pm3\pi$.

To sum up, in a $\widetilde{C}_{N,p}$ symmetric system satisfying $p/N=p'/N'$
with two co-prime numbers $p'$ (an odd integer) and $N'$ (an even integer),
two distinct $\widetilde{C}_{N,p}$ eigenstates should be
connected to each other at the Brillouin zone boundary $k_{z}=\pm N'\pi/2$.
Namely, the eigenstate with the eigenvalue $\widetilde{J}_{m}(k_{z})$
should be smoothly connected to the other eigenstate with the eigenvalue
$\widetilde{J}_{m-N/2}(k_{z})$ at the Brillouin zone boundary
$k_{z}=\pm(N'\pi)/2$ in the following way,
\begin{align}\label{eqn:constraint_general}
\widetilde{J}_{m}(k_{z}+N'\pi)&=\exp\Big[i\pi\frac{(2m-Np'+1)}{N}\Big]\exp\Big(-i\frac{p'}{N'}k_{z}\Big)
\nonumber\\
&=\widetilde{J}_{m-Np'/2}(k_{z})
\nonumber\\
&=\widetilde{J}_{m-N/2}(k_{z}),
\end{align}
where we have used the fact that $p'$ is an odd integer and
$m$ is well-defined modulo $N$.
It is interesting to note that the eigenvalue $\widetilde{J}_{m}(k_{z})$
at the zone boundary $k_{z}=\pm N'\pi/2$ becomes
\begin{align}\label{eqn:screw_BZboundary}
\widetilde{J}_{m}(k_{z}=\pm N'\pi/2)&=J_{m}\exp\Big(\mp i\frac{p'}{2}\pi\Big)
\nonumber\\
&=\mp iJ_{m}(-1)^{(p'-1)/2}.
\end{align}
Namely, the eigenvalue $\widetilde{J}_{m}(k_{z})$ is simply
given by $\pm iJ_{m}$ at the zone boundary.
This additional factor $\pm i$ gives rise to the following
relations between
the screw rotation $\widetilde{C}_{N,p}$ and the inversion $P$
at the zone boundary $\bm{k}_{\pm}=(0,0,\pm N'\pi/2)$,
\begin{align}
P\widetilde{C}_{N,p}\Big|\bm{k}_{\pm}\Big\rangle&=\mp i(-1)^{(p'-1)/2}PC_{N}\Big|\bm{k}_{\pm}\Big\rangle,
\nonumber\\
\widetilde{C}_{N,p}P\Big|\bm{k}_{\pm}\Big\rangle&=\pm i(-1)^{(p'-1)/2}PC_{N}\Big|\bm{k}_{\pm}\Big\rangle,
\end{align}
which can be easily derived from Eq.~(\ref{eqn:PRn2}) and (\ref{eqn:RnP2}).
Hence $P$ and $\widetilde{C}_{N,p}$ anticommute when the Bloch state $\ket{\bm{k}_{\pm}=(0,0,\pm N'\pi/2)}$
is used as a basis for the representation.
From this, we obtain the following general principle to create a stable Dirac semimetal with a single
Dirac point. Namely,
\begin{itemize}
\item  The Dirac point should be located at a TRIM at the Brillouin zone
boundary ($k_{z}=\pm N'\pi/2$) where the screw rotation symmetry anti-commutes
with the inversion symmetry.
\end{itemize}

\begin{figure}[t]
\centering
\includegraphics[width=8.5 cm]{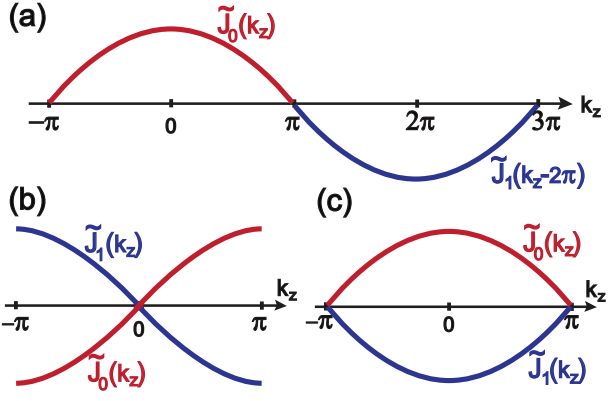}
\caption{
(a) An example of the band connection required by the screw rotation in Eq.~(\ref{eqn:screw21_connection2}).
For each band, the eigenvalue of the screw rotation $\widetilde{C}_{2,1}$
is marked in the figure.
(b,c) Possible band structure protected by
a two-fold screw rotation
$\widetilde{C}_{2,1}$ with a Dirac point at $k_{z}=0$
and at $k_{z}=\pi$, respectively.
}
\label{fig:screw21}
\end{figure}

\subsection{Applications}
In the following, we examine
the possible Dirac semimetals with a single Dirac point
by considering various screw rotation symmetries explicitly.
\subsubsection{Two-fold screw rotation $\widetilde{C}_{2,1}$}

A two-fold screw rotation symmetry $\widetilde{C}_{2,1}$
has the following two eigenvalues
\begin{align}
\widetilde{J}_{0}(k_{z})&=J_{0}\exp\Big(-i\frac{1}{2}k_{z}\Big)=\exp\Big[-i\frac{1}{2}(k_{z}-\pi)\Big],
\nonumber\\
\widetilde{J}_{1}(k_{z})&=J_{1}\exp\Big(-i\frac{1}{2}k_{z}\Big)=\exp\Big[-i\frac{1}{2}(k_{z}-3\pi)\Big],
\end{align}
where
\begin{align}\label{eqn:screw21_connection1}
\widetilde{J}_{m}(k_{z}=-\pi)&\neq\widetilde{J}_{m}(k_{z}=\pi),
\end{align}
for $m=0,1$
and
\begin{align}\label{eqn:screw21_connection2}
\widetilde{J}_{0}(k_{z})&=\widetilde{J}_{1}(k_{z}+2\pi),
\nonumber\\
\widetilde{J}_{1}(k_{z})&=\widetilde{J}_{0}(k_{z}+2\pi).
\end{align}
Now we prepare two bands $\Psi_{0}(k_{z})$ and $\Psi_{1}(k_{z})$ with an eigenvalue $\widetilde{J}_{0}(k_{z})$ and $\widetilde{J}_{1}(k_{z})$, respectively,
and construct a band structure with a Dirac point.
Here the crucial point is that
the band $\Psi_{0}(k_{z})$ ($\Psi_{1}(k_{z})$) should be smoothly connected
to the other band $\Psi_{1}(k_{z})$ ($\Psi_{0}(k_{z})$) at the Brillouin zone boundary ($k_{z}=\pm\pi$)
to satisfy Eq.~(\ref{eqn:screw21_connection1}) and (\ref{eqn:screw21_connection2})
as shown in Fig.~\ref{fig:screw21}.
This naturally gives rise to a band structure with a single band crossing point
at a TRIM.
Considering the $P$ or $T$ symmetry,
there are two possible band structures having a single band crossing point
as shown in Fig.~\ref{fig:screw21} (b) and (c).
In each case, the band crossing point locates at a TRIM either at $k_{z}=0$ (Fig.~\ref{fig:screw21} (b))
or at $k_{z}=\pi$ (Fig.~\ref{fig:screw21} (c)).
However, let us note that, due to the $TP$ symmetry,
the state with the eigenvalue $\widetilde{J}_{0}(k_{z})$ ($\widetilde{J}_{1}(k_{z})$) should be
locally degenerate with the other state with the eigenvalue
$\widetilde{J}_{1}(k_{z})$ ($\widetilde{J}_{0}(k_{z})$) at each momentum $k_{z}$.
This requires that both conduction band and valence band should have the same eigenvalues
of $\widetilde{J}_{0}(k_{z})$ and $\widetilde{J}_{1}(k_{z})$, hence a stable Dirac point cannot be created
due to the finite hybridization between the valence and conduction bands.

\begin{figure}[t]
\centering
\includegraphics[width=8.5 cm]{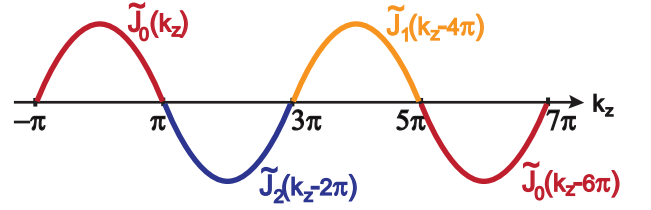}
\caption{
An example of the band connection required by
$\widetilde{C}_{3,1}$ symmetry shown in Eq.(\ref{eqn:screw31_connection2}).
Two bands with the eigenvalues $\widetilde{J}_{0}(k_{z})$ and $\widetilde{J}_{1}(k_{z})$
($\widetilde{J}_{0}(k_{z})$ and $\widetilde{J}_{1}(k_{z})$) form a conduction (valence) band.
However, each degenerate pair violate the $PT$ symmetry.
}
\label{fig:screw31}
\end{figure}

\subsubsection{Three-fold screw rotation}
In the case of a three-fold screw rotation $\widetilde{C}_{3,1}$,
there are three possible eigenvalues given by
\begin{align}
\widetilde{J}_{0}(k_{z})&=J_{0}\exp\Big(-i\frac{1}{3}k_{z}\Big)=\exp\Big[-i\frac{1}{3}(k_{z}-\pi)\Big],
\nonumber\\
\widetilde{J}_{1}(k_{z})&=J_{1}\exp\Big(-i\frac{1}{3}k_{z}\Big)=\exp\Big[-i\frac{1}{3}(k_{z}-3\pi)\Big],
\nonumber\\
\widetilde{J}_{2}(k_{z})&=J_{1}\exp\Big(-i\frac{1}{3}k_{z}\Big)=\exp\Big[-i\frac{1}{3}(k_{z}-5\pi)\Big],
\end{align}
where
\begin{align}\label{eqn:screw31_connection1}
\widetilde{J}_{m}(k_{z}=-\pi)&\neq\widetilde{J}_{m}(k_{z}=\pi),
\end{align}
for $m=0,1,2$ and
\begin{align}\label{eqn:screw31_connection2}
\widetilde{J}_{0}(k_{z})&=\widetilde{J}_{1}(k_{z}+2\pi),
\nonumber\\
\widetilde{J}_{1}(k_{z})&=\widetilde{J}_{2}(k_{z}+2\pi),
\\
\widetilde{J}_{2}(k_{z})&=\widetilde{J}_{0}(k_{z}+2\pi),
\nonumber
\end{align}

Now we prepare three bands $\Psi_{0,1,2}(k_{z})$ with an eigenvalue $\widetilde{J}_{0,1,2}(k_{z})$, respectively,
and construct a band structure with a Dirac point.
To satisfy Eq.~(\ref{eqn:screw31_connection1}) and (\ref{eqn:screw31_connection2}),
each band with a given $\widetilde{C}_{3,1}$ eigenvalue
should be connected to the other two bands with different $\widetilde{C}_{3,1}$ eigenvalues
at each Brillouin zone boundary.
An example of the band connection satisfying Eq.~(\ref{eqn:screw31_connection2})
is shown in Fig.~\ref{fig:screw31}.
According to Fig.~\ref{fig:screw31}, when the band structure is drawn for a reduced Brillouin zone with $k_{z}\in[-\pi,\pi]$,
two bands with the eigenvalues $\widetilde{J}_{0}(k_{z})$ and $\widetilde{J}_{1}(k_{z})$
(or $\widetilde{J}_{0}(k_{z})$ and $\widetilde{J}_{2}(k_{z})$) would form a degenerate band.

However, this band structure is generally incompatible with the $T$ and $P$ symmetries.
The $PT$ symmetry requires the band with an eigenvalue $\widetilde{J}_{0}(k_{z})$
to be locally degenerate with the other band with an eigenvalue $\widetilde{J}_{2}(k_{z})$
whereas the band with an eigenvalue $\widetilde{J}_{1}(k_{z})$
to be locally degenerate with the other band with the same eigenvalue $\widetilde{J}_{1}(k_{z})$.
Namely, the four bands
\begin{align}\label{eqn:basis_screw31}
\{\Psi_{0}(k_{z}),\Psi_{2}(k_{z});\Psi_{1}(k_{z}),\Psi'_{1}(k_{z})\}
\end{align}
would form a basis to create a Dirac point.
In the parenthesis, the first two bands form a conduction (valence) band whereas the last two bands form a valence (conduction) band.
This basis is obviously incompatible with the band connection described in Fig.~\ref{fig:screw31}.
It is straightforward to show that the same problem happens for $\widetilde{C}_{3,2}$ symmetric systems.
Therefore a system with a three-fold screw rotation cannot satisfy the $PT$ symmetry
at the same time, hence cannot have a stable Dirac point.
In fact, every screw symmetric system which does not satisfy
Eq.~(\ref{eqn:p_2pi}) or Eq.~(\ref{eqn:p_2npi}) has
the same problem, thus a Dirac semimetal with a single Dirac point cannot be created.

\subsubsection{Four-fold screw rotation $\widetilde{C}_{4,1}$}
\begin{figure}[t]
\centering
\includegraphics[width=8.5 cm]{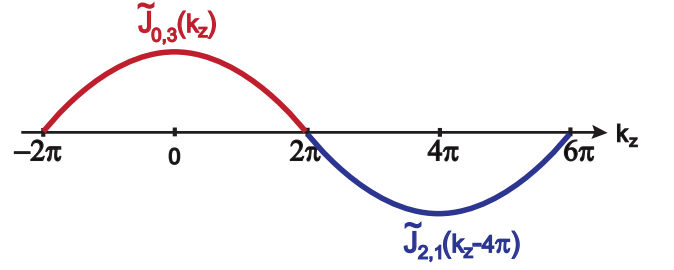}
\caption{
An example of the band connection required by
$\widetilde{C}_{4,1}$ symmetry shown in Eq.(\ref{eqn:screw41_connection4}).
It is assumed that the system is $4\pi$ periodic.
The band with the eigenvalue $\widetilde{J}_{0}(k_{z})$ ($\widetilde{J}_{3}(k_{z})$)
should be connected with the other hand with the eigenvalue $\widetilde{J}_{2}(k_{z})$ ($\widetilde{J}_{1}(k_{z})$)
at the zone boundary.
Moreover, due to the $PT$ symmetry, two bands with the eigenvalues $\widetilde{J}_{0}(k_{z})$ and $\widetilde{J}_{3}(k_{z})$
(or, $\widetilde{J}_{2}(k_{z})$ and $\widetilde{J}_{1}(k_{z})$) should be locally degenerate at each momentum.
In this case, the band connection required by the screw rotation is compatible
with the $PT$ symmetry.
}
\label{fig:screw41}
\end{figure}
In the case of the four-fold screw rotation $\widetilde{C}_{4,1}$,
there are four possible eigenvalues given by
\begin{align}
\widetilde{J}_{0}(k_{z})&=J_{0}\exp\Big(-i\frac{1}{4}k_{z}\Big)=\exp\Big[-i\frac{1}{4}(k_{z}-\pi)\Big],
\nonumber\\
\widetilde{J}_{1}(k_{z})&=J_{1}\exp\Big(-i\frac{1}{4}k_{z}\Big)=\exp\Big[-i\frac{1}{4}(k_{z}-3\pi)\Big],
\nonumber\\
\widetilde{J}_{2}(k_{z})&=J_{2}\exp\Big(-i\frac{1}{4}k_{z}\Big)=\exp\Big[-i\frac{1}{4}(k_{z}-5\pi)\Big],
\nonumber\\
\widetilde{J}_{3}(k_{z})&=J_{3}\exp\Big(-i\frac{1}{4}k_{z}\Big)=\exp\Big[-i\frac{1}{4}(k_{z}-7\pi)\Big].
\end{align}
If the system is $2\pi$ periodic along the $k_{z}$ direction, we obtain
\begin{align}\label{eqn:screw41_connection1}
\widetilde{J}_{m}(k_{z}=-\pi)&\neq\widetilde{J}_{m}(k_{z}=\pi),
\end{align}
for $m=0,1,2,3$
and
\begin{align}\label{eqn:screw41_connection2}
\widetilde{J}_{m}(k_{z})&=\widetilde{J}_{m+1}(k_{z}+2\pi),
\end{align}
hence the state with the eigenvalue $\widetilde{J}_{m}(k_{z})$ should be connected
to the state with the eigenvalue $\widetilde{J}_{m+1}(k_{z})$ at the Brillouin zone boundary ($k_{z}=\pm\pi$).
However, this band connection is not compatible with the $PT$ symmetry of the system,
which requires the state with $\widetilde{J}_{0}(k_{z})$ ($\widetilde{J}_{1}(k_{z})$) to be degenerate with
the state with $\widetilde{J}_{3}(k_{z})$ ($\widetilde{J}_{2}(k_{z})$).

On the other hand, if the system is $4\pi$ periodic along the $k_{z}$ direction,
\begin{align}\label{eqn:screw41_connection3}
\widetilde{J}_{m=0,1,2,3}(k_{z}=-2\pi)&\neq\widetilde{J}_{m=0,1,2,3}(k_{z}=2\pi),
\end{align}
and
\begin{align}\label{eqn:screw41_connection4}
\widetilde{J}_{m}(k_{z})&=\widetilde{J}_{m+2}(k_{z}+4\pi).
\end{align}
This band connection is compatible with the $PT$ symmetry.
The basis for creation of a Dirac semimetal with a single Dirac point at the Brillouin zone boundary
is given by
\begin{align}\label{eqn:screw41_basis}
\{\Psi_{0}(k_{z}),\Psi_{3}(k_{z});\Psi_{1}(k_{z}),\Psi_{2}(k_{z})\}.
\end{align}

\subsubsection{Four-fold screw rotation $\widetilde{C}_{4,2}$}
In the case of the four-fold screw rotation $\widetilde{C}_{4,2}$,
there are four possible eigenvalues given by
\begin{align}
\widetilde{J}_{0}(k_{z})&=J_{0}\exp\Big(-i\frac{1}{2}k_{z}\Big)=\exp\Big[-i\frac{1}{2}(k_{z}-\frac{1}{2}\pi)\Big],
\nonumber\\
\widetilde{J}_{1}(k_{z})&=J_{1}\exp\Big(-i\frac{1}{2}k_{z}\Big)=\exp\Big[-i\frac{1}{2}(k_{z}-\frac{3}{2}\pi)\Big],
\nonumber\\
\widetilde{J}_{2}(k_{z})&=J_{2}\exp\Big(-i\frac{1}{2}k_{z}\Big)=\exp\Big[-i\frac{1}{2}(k_{z}-\frac{5}{2}\pi)\Big],
\nonumber\\
\widetilde{J}_{3}(k_{z})&=J_{3}\exp\Big(-i\frac{1}{2}k_{z}\Big)=\exp\Big[-i\frac{1}{2}(k_{z}-\frac{7}{2}\pi)\Big],
\end{align}
where
\begin{align}\label{eqn:screw42_connection1}
\widetilde{J}_{m}(k_{z}=-\pi)&\neq\widetilde{J}_{m}(k_{z}=\pi),
\end{align}
for $m=0,1,2,3$
and
\begin{align}\label{eqn:screw42_connection2}
\widetilde{J}_{m}(k_{z})&=\widetilde{J}_{m+2}(k_{z}+2\pi),
\nonumber
\end{align}
hence the state with the eigenvalue $\widetilde{J}_{0}(k_{z})$ ($\widetilde{J}_{1}(k_{z})$) should be connected
with the state with the eigenvalue $\widetilde{J}_{2}(k_{z})$ ($\widetilde{J}_{3}(k_{z})$) at the Brillouin zone boundary ($k_{z}=\pm\pi$).
This band connection is compatible with
the $PT$ symmetry.
The basis for creation of a Dirac semimetal with a single Dirac point at the Brillouin zone boundary
is given by
\begin{align}
\{\Psi_{0}(k_{z}),\Psi_{3}(k_{z});\Psi_{1}(k_{z}),\Psi_{2}(k_{z})\}.
\end{align}

\subsubsection{Four-fold screw rotation $\widetilde{C}_{4,3}$}
In the case of the four-fold screw rotation $\widetilde{C}_{4,3}$,
there are four possible eigenvalues given by
\begin{align}
\widetilde{J}_{0}(k_{z})&=J_{0}\exp\Big(-i\frac{3}{4}k_{z}\Big)=\exp\Big[-i\frac{3}{4}(k_{z}-\frac{1}{3}\pi)\Big],
\nonumber\\
\widetilde{J}_{1}(k_{z})&=J_{1}\exp\Big(-i\frac{3}{4}k_{z}\Big)=\exp\Big[-i\frac{3}{4}(k_{z}-\frac{3}{3}\pi)\Big],
\nonumber\\
\widetilde{J}_{2}(k_{z})&=J_{2}\exp\Big(-i\frac{3}{4}k_{z}\Big)=\exp\Big[-i\frac{3}{4}(k_{z}-\frac{5}{3}\pi)\Big],
\nonumber\\
\widetilde{J}_{3}(k_{z})&=J_{3}\exp\Big(-i\frac{3}{4}k_{z}\Big)=\exp\Big[-i\frac{3}{4}(k_{z}-\frac{7}{3}\pi)\Big].
\end{align}
Similar to the case of $\widetilde{C}_{4,1}$ symmetric systems,
a Dirac semimetal with a single Dirac point can be created only if the system is $4\pi$ periodic along
the $k_{z}$ direction.
Then
\begin{align}\label{eqn:screw43_connection1}
\widetilde{J}_{m}(k_{z}=-2\pi)&\neq\widetilde{J}_{m}(k_{z}=2\pi),
\end{align}
for $m=0,1,2,3$
and
\begin{align}\label{eqn:screw43_connection2}
\widetilde{J}_{m}(k_{z})&=\widetilde{J}_{m-2}(k_{z}+4\pi).
\end{align}
The basis for creation of a Dirac semimetal with a single Dirac point at the Brillouin zone boundary
is again
\begin{align}
\{\Psi_{0}(k_{z}),\Psi_{3}(k_{z});\Psi_{1}(k_{z}),\Psi_{2}(k_{z})\}.
\end{align}

\subsubsection{Six-fold screw rotation $\widetilde{C}_{6,3}$}
In the case of the six-fold screw rotation $\widetilde{C}_{6,3}$,
there are six possible eigenvalues given by
\begin{align}
\widetilde{J}_{m}(k_{z})&=J_{m}\exp\Big(-i\frac{1}{2}k_{z}\Big)
\nonumber\\
&=\exp\Big[-i\frac{1}{2}\Big(k_{z}-\frac{2m+1}{3}\pi\Big)\Big],
\end{align}
where $m=1,2,...,6$.
When the system is $2\pi$ periodic along the $k_{z}$ direction,
these eigenvalues satisfy
\begin{align}\label{eqn:screw63_connection1}
\widetilde{J}_{m}(k_{z}=-\pi)&\neq\widetilde{J}_{m}(k_{z}=\pi),
\end{align}
and
\begin{align}\label{eqn:screw63_connection2}
\widetilde{J}_{m}(k_{z})&=\widetilde{J}_{m+3}(k_{z}+2\pi),
\end{align}
hence the state with the eigenvalue $\widetilde{J}_{0,1,2}(k_{z})$ should be connected
to the state with the eigenvalue $\widetilde{J}_{3,4,5}(k_{z})$ at the Brillouin zone boundary ($k_{z}=\pm\pi$),
respectively.
Due to the $T$ and the $P$ symmetries, the pair of states with the eigenvalues
$\{\widetilde{J}_{0}(k_{z}),\widetilde{J}_{5}(k_{z})\}$,
$\{\widetilde{J}_{1}(k_{z}),\widetilde{J}_{4}(k_{z})\}$,
$\{\widetilde{J}_{2}(k_{z}),\widetilde{J}_{3}(k_{z})\}$ should
be locally degenerate at each momentum $k_{z}$.
Considering the band connection described in Eq.~(\ref{eqn:screw63_connection2}),
we find the following basis
\begin{align}\label{eqn:C63_basis}
\{\Psi_{0}(k_{z}),\Psi_{5}(k_{z});\Psi_{3}(k_{z}),\Psi_{2}(k_{z})\},
\end{align}
which can create a Dirac semimetal
with a single Dirac point at the Brillouin zone boundary.

\subsubsection{Six-fold screw rotation $\widetilde{C}_{6,p\neq3}$}
In the case of the six-fold screw rotation $\widetilde{C}_{6,p\neq3}$,
a single Dirac point can be created only if the system is $6\pi$ periodic
along the $k_{z}$ direction.

In $\widetilde{C}_{6,1}$ symmetric systems,
there are six possible eigenvalues given by
\begin{align}
\widetilde{J}_{m}(k_{z})&=J_{m}\exp\Big(-i\frac{1}{6}k_{z}\Big)
\nonumber\\
&=\exp\Big\{-i\frac{1}{6}[k_{z}-(2m+1)\pi]\Big\},
\end{align}
where $m=1,2,...,6$.
These eigenvalues satisfy
\begin{align}\label{eqn:screw61_connection1}
\widetilde{J}_{m}(k_{z}=-3\pi)&\neq\widetilde{J}_{m}(k_{z}=3\pi),
\end{align}
and
\begin{align}\label{eqn:screw61_connection2}
\widetilde{J}_{m}(k_{z})&=\widetilde{J}_{m+3}(k_{z}+6\pi),
\end{align}

In $\widetilde{C}_{6,5}$ symmetric systems
there are six possible eigenvalues given by
\begin{align}
\widetilde{J}_{m}(k_{z})&=J_{m}\exp\Big(-i\frac{5}{6}k_{z}\Big)
\nonumber\\
&=\exp\Big[-i\frac{5}{6}\Big(k_{z}-\frac{2m+1}{5}\pi\Big)\Big],
\end{align}
where $m=1,2,...,6$.
These eigenvalues satisfy
\begin{align}\label{eqn:screw61_connection1}
\widetilde{J}_{m}(k_{z}=-3\pi)&\neq\widetilde{J}_{m}(k_{z}=3\pi),
\end{align}
and
\begin{align}\label{eqn:screw61_connection2}
\widetilde{J}_{m}(k_{z})&=\widetilde{J}_{m+3}(k_{z}+6\pi).
\end{align}
In both $\widetilde{C}_{6,1}$ and $\widetilde{C}_{6,5}$ symmetric cases, we find the following basis
\begin{align}
\{\Psi_{0}(k_{z}),\Psi_{5}(k_{z});\Psi_{3}(k_{z}),\Psi_{2}(k_{z})\},
\end{align}
which can create a Dirac semimetal
with a single Dirac point at the Brillouin zone boundary.

Finally, it is straightforward to show that
$\widetilde{C}_{6,2}$ and $\widetilde{C}_{6,4}$ symmetric cases,
which are similar to the system with a three-fold screw symmetry,
cannot support a Dirac semimetal with a single Dirac point.

\section{\label{sec:class2} Class II Dirac semimetals}
Based on the discussion in Sec.~\ref{sec:class2_screwrotation}, we define
class II Dirac semimetals in the following way.
Class II Dirac semimetals are associated with a special type of rotation symmetry $\widetilde{C}_{N}$ which
anti-commutes with the inversion symmetry, i.e.,
\begin{align}
\{P, \widetilde{C}_N\}=0.
\end{align}
Considering that i) a rotation operator $C$ generally has a form
of $C=\exp(i\theta J)$ with the angular momentum operator $J$ and the rotation angle $\theta$,
and ii) $J$ is a pseudovector which is even under the inversion symmetry $P$, i.e.,
$PJP^{-1}=J$,
the anti-commutation relation between $P$ and $\widetilde{C}_{N}$
looks quite unusual.
However, as discussed in the previous section,
non-symmorphic screw rotation symmetries can generally satisfy
such anti-commutation relations at the Brillouin zone boundary~\cite{Sigrist, nonsymmorphicTI},
when Bloch states are used as a basis for the representation of $P$ and $\widetilde{C}_{N}$.
We first describe the physical consequence resulting from the relation
$\{P, \widetilde{C}_N\}=0$ and the topological charge of the associated Dirac point. After that
we describe how the Dirac point protected by
a screw rotation symmetry leads to a class II Dirac semimetal, and its associated topological charge
can be determined.

\subsection{Symmetry constraint and band connection}
Let us consider the constraints to the rotation eigenvalues $\widetilde{J}_{m}$
due to the symmetries $P$, $T$, $\widetilde{C}_{N}$ satisfying
\begin{align}
&[T,\widetilde{C}_N]=0,\quad \{P, \widetilde{C}_N\}=0,\quad [T,P]=0,
\label{eqn:conditions for class II}
\end{align}
\begin{align}
&T^{2}=-1,\quad P^{2}=1.
\end{align}
Moreover, considering Eq.~(\ref{eqn:screw_BZboundary}), we assume
the $\widetilde{C}_{N}$ eigenvalue $\widetilde{J}_{m}$ to have the following form,
\begin{align}
\widetilde{J}_{m}=iJ_{m},
\end{align}
where $J_{m}=\exp[i\pi(2m+1)/N]$.
The action of the $PT$ on an eigenvector $\ket{\psi_{m}(\bm{k})}$ of $\widetilde{C}_N$
with an eigenvalue $\widetilde{J}_{m}$ gives
\begin{align}\label{eqn:constraint_TP}
\widetilde{C}_N PT \ket{\psi_{m}(\bm{k})} &= - PT \widetilde{C}_N \ket{\psi_{m}(\bm{k})} \n
&= - PT \widetilde{J}_m \ket{\psi_{m}(\bm{k})} \n
&= - \widetilde{J}_m^* PT \ket{\psi_{m}(\bm{k})} \n
&= \widetilde{J}_{N-m-1} PT \ket{\psi_{m}(\bm{k})}\n
&\equiv \widetilde{J}_{N-m-1} \ket{\psi_{N-m-1}(\bm{k})},
\end{align}
from which we find $PT\ket{\psi_m}$ is a degenerate eigenvector with the eigenvalue
$\widetilde{J}_{N-m-1}$.
On the other hand, $P$ and $T$ transform the eigenvector $\ket{\psi_m}$
at $\bm{k}$ to another eigenvector at $-\bm{k}$ as
\begin{align}\label{eqn:constraint_T}
\widetilde{C}_N T \ket{\psi_{m}(\bm{k})} &= T \widetilde{C}_N \ket{\psi_{m}(\bm{k})} \n
&= T \widetilde{J}_m \ket{\psi_{m}(\bm{k})} \n
&= \widetilde{J}_m^* T \ket{\psi_{m}(\bm{k})} \n
&= \widetilde{J}_{\frac{N}{2}-m-1} T \ket{\psi_{m}(\bm{k})} \n
&\equiv \widetilde{J}_{\frac{N}{2}-m-1} \ket{\psi_{\frac{N}{2}-m-1}(-\bm{k})},
\end{align}
and
\begin{align}\label{eqn:constraint_P}
\widetilde{C}_N P \ket{\psi_{m}(\bm{k})} &= - P \widetilde{C}_N \ket{\psi_{m}(\bm{k})} \n
&= - P \widetilde{J}_m \ket{\psi_{m}(\bm{k})} \n
&= - \widetilde{J}_m P \ket{\psi_{m}(\bm{k})} \n
&= \widetilde{J}_{\frac{N}{2}+m} P \ket{\psi_{m}(\bm{k})} \n
&\equiv \widetilde{J}_{\frac{N}{2}+m} \ket{\psi_{\frac{N}{2}+m}(-\bm{k})}.
\end{align}
Since $\frac{N}{2}-m-1$ and $\frac{N}{2}+m$ should be an integer, we find that $N$ should be an even number.
Hence the class II Dirac semimetal cannot exist in systems with $\widetilde{C}_{3}$ symmetry.
From Eqs.~(\ref{eqn:constraint_TP}),
~(\ref{eqn:constraint_T}), and~(\ref{eqn:constraint_P}),
we find that if $\ket{\psi_{m}(\bm{k})}$
is an eigenstate at $\bm{k}$, $\ket{\psi_{N-m-1}(\bm{k})}$ is also a degenerate
eigenstate at the same momentum $\bm{k}$,
whereas $\ket{\psi_{\frac{N}{2}-m-1}(-\bm{k})}$ and $\ket{\psi_{\frac{N}{2}+m}(-\bm{k})}$
are degenerate at $-\bm{k}$ with the same energy as $\ket{\psi_{m}(\bm{k})}$.
From this, we can infer the band connection near a TRIM ($\bm{k}_{\text{TRIM}}$) where $\bm{k}$ and $-\bm{k}$ are equivalent.
The question is whether two doubly-degenerate states at $\bm{k}$ and $-\bm{k}$
are crossing or smoothly connected at $\bm{k}_{\text{TRIM}}$.
A smooth connection of degenerate bands requires the rotation eigenvalues
of the states at $\bm{k}$ and $-\bm{k}$ to be identical,
which is obviously not satisfied in this case.
Therefore there should be a band crossing at $\bm{k}_{\text{TRIM}}$
where the Dirac point locates.

\begin{figure}[t]
\centering
\includegraphics[width=8.5 cm]{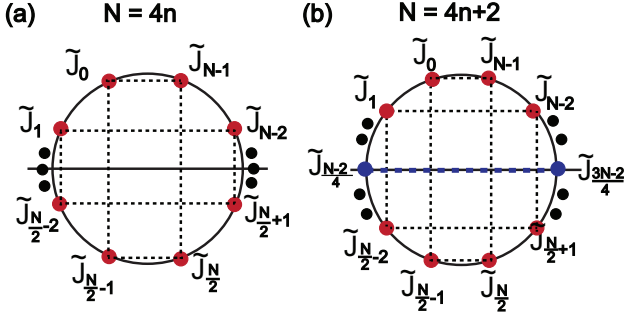}
\caption{
Constraints on the rotation eigenvalues (a) for $N=4n$, (b) for $N=4n+2$
with an integer $n$.
The (black) solid circle indicates a unit circle in the complex plane,
and each (red) dot on the circle denotes $\widetilde{J}_{m}$.
Four dots connected by a dotted line are related by the $P$, $T$, and $PT$ symmetries,
hence only one of them is independent.
In the case of (b), the topological charge associated with
$\widetilde{J}_{(N-2)/4}$ and $\widetilde{J}_{(3N-2)/4}$ is zero.
}
\label{fig:anticommuting}
\end{figure}\label{table:3}

\subsection{Topological charge \label{subsec:class2 charge}}
From Eqs.~(\ref{eqn:constraint_TP}),
~(\ref{eqn:constraint_T}), and~(\ref{eqn:constraint_P}),
we can easily find the constraints to
the topological charge $\nu_{m}$ of the Dirac point locating
at a TRIM. At first, the $PT$ symmetry requires that
\begin{align}
\nu_{m}&=\nu_{N-m-1}
\nonumber\\
\nu_{\frac{N}{2}-m-1}&=\nu_{\frac{N}{2}+m}.
\end{align}

Moreover, since $P$ or $T$ symmetry
interchanges the north and south pole surrounding
a Dirac point at a TRIM, we find
\begin{align}\label{eqn:constraint_anticommuting}
\nu_{m}&=\nu_{N-m-1}=
-\nu_{\frac{N}{2}-m-1}=-\nu_{\frac{N}{2}+m}.
\end{align}
This constraint reduces the number of independent topological numbers $\nu_{m}$.
Considering that $N$ is an even integer, we distinguish two cases, i.e., when $N=4n$
and when $N=4n+2$ with an integer $n$.
In each case, the relation between different eigenstates
are described in Fig.~\ref{fig:anticommuting},
from which we find the topological invariant
\begin{align}
(\nu_{0},...,\nu_{\frac{N}{4}-1})\in\mathbb{Z}^{\frac{N}{4}}\quad &\text{if}\quad N=4n,
\nonumber\\
(\nu_{0},...,\nu_{\frac{N-6}{4}})\in\mathbb{Z}^{\frac{N-2}{4}}\quad &\text{if}\quad  N=4n+2.
\end{align}
Hence the topological charge of the system with $\widetilde{C}_{4}$ or $\widetilde{C}_{6}$ symmetry is an element
of $\mathbb{Z}$ whereas $\widetilde{C}_{2}$ or $\widetilde{C}_{3}$ symmetric system
cannot support a Dirac point with a nonzero topological charge.

\begin{table}[h]
\begin{tabular}{c c c}
\hline
\hline
$\widetilde{C}_{N}$ & & Topological charge \\
\hline
\hline
$\widetilde{C}_{2}$ & &  Not allowed\\
$\widetilde{C}_{3}$ & &  Not allowed\\
$\widetilde{C}_{4}$ & & $\mathbb{Z}$ \\
$\widetilde{C}_{6}$ & & $\mathbb{Z}$  \\
\hline \hline
\end{tabular}
\caption{Summary of topological charges of class II Dirac semimetals.}
\end{table}\label{table:4}

\subsection{Applications: classification of stable Dirac points in 4-band systems}
Due to the $PT$ symmetry,
$\{|\psi_{m}(\bm{k})\rangle, |\psi_{N-m-1}(\bm{k})\rangle\}$
and $\{|\psi_{N/2-m-1}(\bm{k})\rangle, |\psi_{N/2+m}(\bm{k})\rangle\}$ form degenerate
pairs at each momentum $\bm{k}$, and these four states cross
at a TRIM, and create a Dirac point.
A 4-band model can be constructed by using these four states.

\subsubsection{$\widetilde{C}_{2}$ symmetric systems}

Possible $\widetilde{J}_{m}$ values are $\widetilde{J}_{m=0}=i\exp(i\frac{1}{2}\pi)$
and $\widetilde{J}_{m=1}=i\exp(i\frac{3}{2}\pi)$.
Due to the $PT$ symmetry, $\{|\psi_{m=0}(\bm{k})\rangle, |\psi_{m=1}(\bm{k})\rangle\}$
and $\{|\psi'_{m=0}(\bm{k})\rangle, |\psi'_{m=1}(\bm{k})\rangle\}$ form degenerate pairs.
The symmetry constraint in Eq.~(\ref{eqn:constraint_anticommuting})
requires
\begin{eqnarray}
\nu_{m=0}=\nu_{m=1}=0.
\end{eqnarray}
Hence a $\widetilde{C}_{2}$ invariant system cannot support a stable Dirac point at a TRIM.

\subsubsection{$\widetilde{C}_{3}$ symmetric systems}

Possible $\widetilde{J}_{m}$ values are $\widetilde{J}_{m=0}=i\exp(i\frac{1}{3}\pi)$,
$\widetilde{J}_{m=1}=i\exp(i\pi)=-i$,
and $\widetilde{J}_{m=2}=i\exp(i\frac{5}{3}\pi)$.
In the case of $\ket{\psi_{m=1}}$, the $P$ symmetry requires that
$\widetilde{C}_{3} P \ket{\psi_{m=1}} = - P \widetilde{C}_{3} \ket{\psi_{m=1}}= iP\ket{\psi_{m=1}}$.
Thus $P \ket{\psi_{m=1}}$ should be an eigenstate of $\widetilde{C}_{3}$ with the eigenvalue $+i$,
which is not allowed.
Hence a $\widetilde{C}_{3}$ invariant system cannot support a stable Dirac point at a TRIM.

\subsubsection{$\widetilde{C}_{4}$ symmetric systems}

Due to the $PT$ symmetry, $\{|\psi_{m=0}\rangle, |\psi_{m=3}\rangle\}$
and $\{|\psi_{m=1}\rangle, |\psi_{m=2}\rangle\}$ form degenerate pairs.
The constraint in Eq.~(\ref{eqn:constraint_anticommuting})
requires that
\begin{eqnarray}
\overline{\nu}_{1}\equiv \nu_{m=0}=\nu_{m=3}=-\nu_{m=1}=-\nu_{m=2},
\end{eqnarray}
thus
there is only one independent topological invariant
$\overline{\nu}_{1}=\pm1$.

\subsubsection{$\widetilde{C}_{6}$ symmetric systems}

Due to the $P$ and $T$ symmetries, we find a basis
$\{|\psi_{m=0}\rangle, |\psi_{m=5}\rangle;|\psi_{m=2}\rangle, |\psi_{m=3}\rangle\}$.
The constraint in Eq.~(\ref{eqn:constraint_anticommuting}) requires
\begin{align}
\overline{\nu}_{1}&\equiv \nu_{m=0}=\nu_{m=5}=-\nu_{m=2}=-\nu_{m=3},
\end{align}
thus
there is only one independent topological invariant
$\overline{\nu}_{1}=\pm1$.

\begin{table}[h]
\begin{tabular}{c c c c c}
\hline
\hline
$\widetilde{C}_{N}$  & & 4-band model & & Materials \\
\hline
\hline
$\widetilde{C}_{2}$  & & Not allowed & & \\
$\widetilde{C}_{3}$ & & Not allowed & &\\
$\widetilde{C}_{4}$ & & $\overline{\nu}_{1}=\pm1$ & & $\beta$-BiO$_{2}$~[\onlinecite{Young1}]\\
$\widetilde{C}_{6}$ & & $\overline{\nu}_{1}=\pm1$& & \\
\hline \hline
\end{tabular}
\caption{Topological charges of class II Dirac semimetals
for 4-band models.}
\end{table}

\subsection{Topological charge of Dirac points protected by screw rotations}
Here we show that the Dirac semimetals protected by screw rotations
belong to the class II, thus the topological charge of
the relevant Dirac point can be determined by following the prescription described in Sec.~\ref{subsec:class2 charge}.
In particular, we resolve a subtle issue associated with the multi-valued nature
of the screw rotation eigenvalues, which we encounter
when we define the topological charge of the Dirac point at the Brillouin zone boundary.
To understand this, let us again introduce a sphere in the momentum space surrounding the Dirac point,
and consider the two points $\bm{k}_{N}$ and $\bm{k}_{S}$ on the sphere passing the rotation axis.
To compare the zero-dimensional topological numbers at these two points,
we need a single-valued wave function which varies smoothly around the Dirac point.
However, since the eigenvalue of a screw rotation $\widetilde{J}_{m}(k_{z})$ is multi-valued,
the relevant eigenstates also change discontinuously at the Brillouin zone boundary.
To remedy this problem, we propose a way to construct a smooth function which is single-valued around the Dirac point
by modifying eigenvectors of $\widetilde{C}_{N,p}$.
Moreover, we show that such a smooth single-valued function
satisfies the algebraic relations shown in Eq.~(\ref{eqn:constraint_TP}),
(\ref{eqn:constraint_T}), and (\ref{eqn:constraint_P}), thus
we prove that a Dirac semimetal protected by a screw rotation
belongs to the class II.

First, we suppose that $p/N=p'/N'$ holds with an even integer $N'$ and an odd integer $p'$ that are coprime,
and the Brillouin zone boundary is located at $k_{z}=N'\pi/2$.
To construct a smooth single-valued wave function
around the Brillouin zone boundary at $k_{z}=N'\pi/2$, we prepare two eigenstates
$\ket{\Psi_{m}(k_{z})}$ and $\ket{\Psi_{m-\frac{N'}{2}p}(k_{z})}$
with $\widetilde{C}_{N,p}$ eigenvalues
$\widetilde{J}_{m}(k_{z})$ and $\widetilde{J}_{m-\frac{N'}{2}p}(k_{z})$, respectively.
Then we define a hybrid wave function around the zone boundary as
\begin{align}
\ket{\widetilde{\psi}_{m}(\delta k_z)} &=
\begin{cases}
\ket{\Psi_{m}(N'\pi/2 + \delta k_z )},  & (\delta k_z \le 0 ) \\
\ket{\Psi_{m-\frac{N'}{2}p}(-N'\pi/2 + \delta k_z )},  & (\delta k_z >0) \\
\end{cases}
\end{align}
which is smooth and single-valued around the zone boundary at $\delta k_z=0$; see Eq.~(\ref{eqn:constraint_general}).
It is straightforward to show that
$\ket{\widetilde{\psi}_{m}(\delta k_z)}$ is also an eigenvector of $\widetilde{C}_{N,p}$ satisfying
\begin{align}
\widetilde{C}_{N,p} \ket{\widetilde \psi_{m}(\delta k_z)} &=
J'_m(\delta k_z) \ket{\widetilde \psi_{m}(\delta k_z )},
\end{align}
where
\begin{align}
J'_m(\delta k_z) &= \exp \left[ i \left(\frac{2m+1}{N} -\frac{p'}{2} \right)\pi\right]
\exp\left( -i \frac{p'}{N'} \delta k_z \right).
\end{align}

Now we determine the $\widetilde{C}_{N,p}$ eigenvalue of $T\ket{\widetilde \psi_{m}(\delta k_z)}$.
At first, if $\delta k_z \le 0$,
we obtain
\begin{align}\label{eqn:screwT1}
T \ket{\widetilde \psi_{m}(\delta k_z)}
&= T \ket{\Psi_{m}(N'\pi/2 + \delta k_z )}\n
&\propto \ket{\Psi_{N-m-1}(-N'\pi/2 - \delta k_z )} \n
&=  \ket{\widetilde{\psi}_{N-m-1+\frac{N'}{2}p}(- \delta k_z )},
\end{align}
where we have used the following relation,
\begin{align}
J^{*}_{m}=J_{N-m-1}.
\end{align}
On the other hand, if $\delta k_z > 0$,
\begin{align}\label{eqn:screwT2}
T \ket{\widetilde \psi_{m}(\delta k_z)}
&= T \ket{\Psi_{m-\frac{N'}{2}p}(-N'\pi/2 + \delta k_z )} \n
&\propto \ket{\Psi_{N-m-1+\frac{N'}{2}p}(N'\pi/2 - \delta k_z )} \n
&=  \ket{\widetilde{\psi}_{N-m-1+\frac{N'}{2}p}(- \delta k_z )}.
\end{align}
From Eqs.~(\ref{eqn:screwT1}) and (\ref{eqn:screwT2}), we obtain
\begin{align}\label{eqn:screwT3}
\widetilde{C}_{N,p} T\ket{\widetilde \psi_{m}(\delta k_z)} &=
J'_{N-m-1+\frac{N'}{2}p}(-\delta k_z) T\ket{\widetilde \psi_{m}(\delta k_z )}\n
&=
J'_{N-m-1+\frac{N}{2}}(-\delta k_z) T\ket{\widetilde \psi_{m}(\delta k_z )},\n
&=
J'_{\frac{N}{2}-m-1}(-\delta k_z) T\ket{\widetilde \psi_{m}(\delta k_z )},
\end{align}
where we have used
\begin{align}
\frac{N'}{2}p=\frac{N}{2}p'=\frac{N}{2} \quad(\textrm{mod}~N),
\end{align}
in which $p'$ is an odd integer.

The $\widetilde{C}_{N,p}$ eigenvalue of $P\ket{\widetilde \psi_{m}(\delta k_z)}$
can also be obtained similarly from the following two relations,
\begin{align}\label{eqn:screwP1}
P \ket{\widetilde \psi_{m}(\delta k_z\leq 0)}
&= P \ket{\Psi_{m}(N'\pi/2 + \delta k_z )}\n
&\propto \ket{\Psi_{m}(-N'\pi/2 - \delta k_z )} \n
&=  \ket{\widetilde{\psi}_{m+\frac{N'}{2}p}(- \delta k_z )}\n
&=  \ket{\widetilde{\psi}_{m+\frac{N}{2}}(- \delta k_z )},
\end{align}
and
\begin{align}\label{eqn:screwP2}
P \ket{\widetilde \psi_{m}(\delta k_z >0)}
&= P \ket{\Psi_{m-\frac{N'}{2}p}(-N'\pi/2 + \delta k_z )} \n
&\propto \ket{\Psi_{m-\frac{N'}{2}p}(N'\pi/2 - \delta k_z )} \n
&=  \ket{\widetilde{\psi}_{m-\frac{N'}{2}p}(- \delta k_z )}\n
&=  \ket{\widetilde{\psi}_{m+\frac{N}{2}}(- \delta k_z )},
\end{align}
where we have used the fact that $m$ is well-defined modulo $N$.
From Eqs.~(\ref{eqn:screwP1}) and (\ref{eqn:screwP2}), we obtain
\begin{align}\label{eqn:screwP3}
\widetilde{C}_{N,p} P\ket{\widetilde \psi_{m}(\delta k_z)} &=
J'_{m+\frac{N}{2}}(-\delta k_z) P\ket{\widetilde \psi_{m}(\delta k_z )}.
\end{align}

Finally,
the $\widetilde{C}_{N,p}$ eigenvalue of $PT\ket{\widetilde \psi_{m}(\delta k_z)}$
can also be obtained by following similar steps, which give
\begin{align}\label{eqn:screwPT3}
\widetilde{C}_{N,p} PT\ket{\widetilde \psi_{m}(\delta k_z)} &=
J'_{N-m-1}(\delta k_z) PT\ket{\widetilde \psi_{m}(\delta k_z )}.
\end{align}
These transformation laws should be compared with Eqs.~(\ref{eqn:constraint_TP}),
(\ref{eqn:constraint_T}), and (\ref{eqn:constraint_P}),
which differ from Eqs.~(\ref{eqn:screwT3}), (\ref{eqn:screwP3}), and (\ref{eqn:screwPT3})
merely due to the momentum-dependent phase factor in $J'_m(\delta k_z)$.
However, since the wave function $\ket{\widetilde \psi_{m}(\delta k_z)}$ and its associated the eigenvalue $J'_m(\delta k_z)$
are smooth and single-valued around $\delta k_z=0$, i.e.,  near the Brillouin zone boundary,
the topological charge of the Dirac point can be defined
by using $J'_m(\delta k_z=0)$ at the two momenta $\bm{k}_{N}$ and $\bm{k}_{S}$ near the Dirac point.
Once $\delta k_z$ is fixed to be $\delta k_z=0$,
Eqs.~(\ref{eqn:constraint_TP}),
(\ref{eqn:constraint_T}), and (\ref{eqn:constraint_P})
are identical to Eqs.~(\ref{eqn:screwT3}), (\ref{eqn:screwP3}), and (\ref{eqn:screwPT3}),
which shows that the Dirac point at the Brillouin zone boundary protected by a screw rotation
belongs to the class II.

\subsection{\label{sec:hcp} Example 1: a class II Dirac semimetal on a hcp lattice}
\begin{figure}[t]
\centering
\includegraphics[width=8.5 cm]{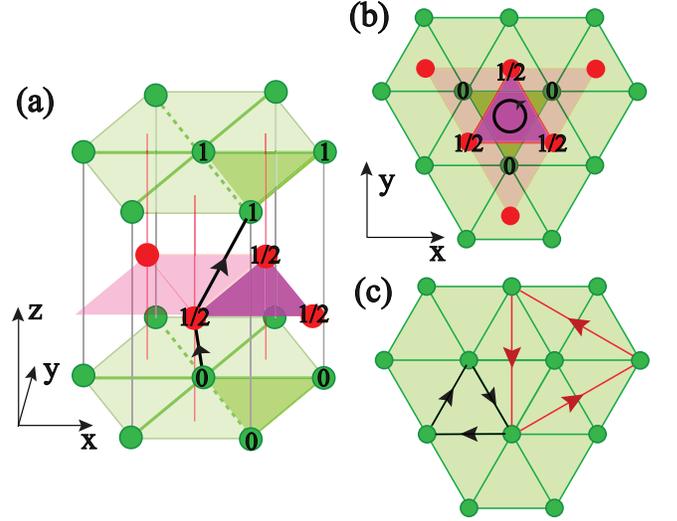}
\caption{
(a) Hexagonal close-packed (hcp) lattice structure and 6-fold screw rotation.
Two sublattice sites are marked with different colors.
The arrows indicate a 6-fold screw rotation about the $z$ axis ($\widetilde{C}_{6,3}$).
The number on a lattice site symbol indicates its $z$-coordinate in the unit of the vertical lattice spacing $c$.
(b) Projection of the lattice to the $xy$ plane.
(c) Spin-dependent complex hopping process between the nearest neighbor sites $(\lambda^{(1)}_{\text{SO}})$
and the next nearest neighbor sites $(\lambda^{(2)}_{\text{SO}})$.
When a spin-up electron on the $A$ sublattice hops
parallel (anti-parallel) to the arrow direction, the corresponding $\nu_{ij}=+1$ ($\nu_{ij}=-1$).
}
\label{fig:hcpscrew}
\end{figure}
\begin{figure}[t]
\centering
\includegraphics[width=8.5 cm]{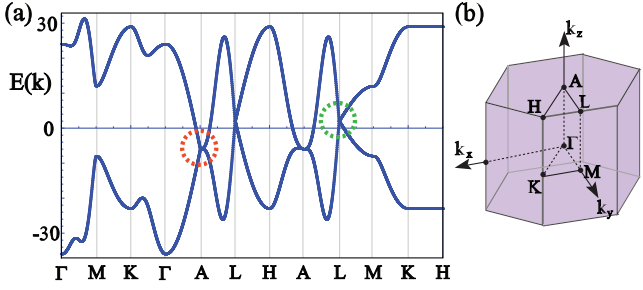}
\caption{
(a) Band structure of the tight binding model
on a hcp lattice.
There are two Dirac points at the $A$ and $L$ points.
(b) First Brillouin zone of the hcp lattice.
}
\label{fig:hcp_dispersion}
\end{figure}

To illustrate the role of screw rotations on the protection of
a Dirac point, let us consider a tight-binding Hamiltonian
on a hexagonal close-packed (hcp) lattice,
which corresponds to the space group $P6_{3}/mmc$ (no. 194).
The hcp lattice is generated by the primitive lattice vectors
$\bm{a}_{1}=a\hat{x}$, $\bm{a}_{2}=a(\frac{1}{2}\hat{x}+\frac{\sqrt{3}}{2}\hat{y})$,
$\bm{a}_{3}=c\hat{z}$, and two sites in a unit cell located at
$\bm{r}_{1}=\bm{0}$ and $\bm{r}_{2}=\frac{1}{3}\bm{a}_{1}+\frac{1}{3}\bm{a}_{2}+\frac{1}{2}\bm{a}_{3}$, respectively.
The crystal has a 6-fold screw rotation symmetry $\widetilde{C}_{6,3}$
about the $z$ axis centered at $\frac{2}{3}\bm{a}_{1}+\frac{2}{3}\bm{a}_{2}$ accompanied
by a partial translation $\frac{1}{2}\bm{a}_{3}$ as shown in Fig.~\ref{fig:hcpscrew}.
To confirm the presence of a single Dirac point at the zone boundary, corresponding to the $A$ point in Fig.~\ref{fig:hcp_dispersion} (b),
we construct the following tight binding model,
\begin{align}
H=&-t_{1}\sum_{\langle ij \rangle}c_{i}^{\dag}c_{j}
-t_{2}\sum_{\langle ij \rangle}c_{i}^{\dag}\tau_{x}c_{j}
\nonumber\\
&+i\lambda^{(1)}_{\text{SO}}\sum_{\langle ij \rangle}\nu^{(1)}_{ij}c_{i}^{\dag}\sigma_{z}\tau_{z}c_{j}
+i\lambda^{(2)}_{\text{SO}}\sum_{\langle\langle ij \rangle\rangle}\nu^{(2)}_{ij}c_{i}^{\dag}\sigma_{z}\tau_{z}c_{j},
\end{align}
where $t_{1}$ ($t_{2}$) indicates the nearest-neighbor hopping
between the same (different) sublattice sites
and $\lambda^{(1)}_{\text{SO}}$ ($\lambda^{(2)}_{\text{SO}}$) denotes
the spin-orbit interaction between the nearest-neighbor (next-nearest-neighbor)
sites. $\nu^{(1,2)}_{ij}=-\nu^{(1,2)}_{ji}$ is $+1$ $(-1)$ if the bond $ij$
is parallel (anti-parallel) to the arrow direction on the bond as shown in Fig.~\ref{fig:hcpscrew} (c).
$\sigma_{x,y,z}$ ($\tau_{x,y,z}$)
are Pauli matrices indicating the spin (sublattice) degrees of freedom.

In the momentum space, the Hamiltonian becomes
\begin{displaymath}
H(\textbf{k})=F_{0}+
\left[ \begin{array}{cc}
(F^{(1)}_{\text{SO}}+F^{(2)}_{\text{SO}})\sigma_{z} & F_{1}\textbf{1}_{2} \\
F_{1}^{*}\textbf{1}_{2} & -(F^{(1)}_{\text{SO}}+F^{(2)}_{\text{SO}})\sigma_{z}
\end{array} \right],
\end{displaymath}
where $\textbf{1}_{2}$ indicates a $2\times 2$ identity matrix
and $F_{0,1}$ and $F^{(1,2)}_{\text{SO}}$ are given by
\begin{align}
F_{0}&=-2t_{1}\left\{\cos(\bm{k}\cdot\bm{a}_{1})+\cos(\bm{k}\cdot\bm{a}_{2})+\cos[\bm{k}\cdot(\bm{a}_{1}-\bm{a}_{2})]\right\},
\nonumber\\
F_{1}&=-2t_{2}\cos\left(\frac{ck_{z}}{2}\right)\left(e^{i\bm{k}\cdot\bm{b}_{1}}+e^{i\bm{k}\cdot\bm{b}_{2}}+e^{i\bm{k}\cdot\bm{b}_{3}}\right),
\nonumber\\
F^{(1)}_{\text{SO}}&=2\lambda^{(1)}_{\text{SO}}\left\{\sin(\bm{k}\cdot\bm{a}_{1})-\sin(\bm{k}\cdot\bm{a}_{2})-\sin[\bm{k}\cdot(\bm{a}_{1}-\bm{a}_{2})]\right\},
\nonumber\\
F^{(2)}_{\text{SO}}&=-2\lambda^{(2)}_{\text{SO}}\left[\sin(3\bm{k}\cdot\bm{b}_{1})+\sin(3\bm{k}\cdot\bm{b}_{2})+\sin(3\bm{k}\cdot\bm{b}_{3})\right],
\end{align}
where $\bm{b}_{1}=\frac{a}{2}\hat{x}+\frac{a}{2\sqrt{3}}\hat{y}$,
$\bm{b}_{2}=-\frac{a}{2}\hat{x}+\frac{a}{2\sqrt{3}}\hat{y}$,
and $\bm{b}_{3}=-\frac{a}{\sqrt{3}}\hat{y}$.
We choose $t_{1}=1$, $t_{2}=5$, and $\lambda^{(1)}_{\text{SO}}=\lambda^{(2)}_{\text{SO}}=5$.
The resulting band structure is shown in Fig.~\ref{fig:hcp_dispersion}.
We can clearly see two Dirac points at $A$ and $L$, respectively.
Between these two points, the Dirac point at $A$ is
the one protected by the 6-fold screw rotation, hence is located
at the boundary of the rotation axis ($z$-axis).

To confirm the symmetry protection of the Dirac point at $A$ and its characteristic dispersion,
let us examine the symmetry of the Hamiltonian.
The symmetries, which are important for the protection of Dirac points, are
the time-reversal $T$, the inversion $P$,
the 6-fold screw rotation $\widetilde{C}_{6,3}=\{C_{6}|\frac{c}{2}\hat{z}\}$, and
the glide symmetry $\widetilde{M}_{y}=\{M_{y}|\frac{c}{2}\hat{z}\}$
where $M_{y}$ transforms the spatial
coordinate $(x,y,z)$ to $(x,-y,z)$.
To find the matrix representation of each symmetry operator,
we can use the following information. At first, for $\bm{k}\rightarrow -\bm{k}$,
we find
\begin{align}
F_{0}(-\bm{k})&=F_{0}(\bm{k}),
\nonumber\\
\text{Re}F_{1}(-\bm{k})&=\text{Re}F_{1}(\bm{k}),
\nonumber\\
\text{Im}F_{1}(-\bm{k})&=-\text{Im}F_{1}(\bm{k}),
\nonumber\\
F^{(1,2)}_{\text{SO}}(-\bm{k})&=-F^{(1,2)}_{\text{SO}}(\bm{k}),
\end{align}
which gives
\begin{align}
P=\tau_{x},\quad T=i\sigma_{y}K,
\end{align}
where $K$ is a complex conjugation operator.
Moreover, under $\pi/3$ rotation about the $z$ axis, we obtain
\begin{align}
(k_{x}+ik_{y})&\rightarrow (k'_{x}+ik'_{y})=(k_{x}+ik_{y})\exp\left(i\frac{\pi}{3}\right),
\nonumber\\
k_{z}&\rightarrow k'_{z}=k_{z}
\end{align}
and
\begin{align}
F_{0}(\bm{k}')&=F_{0}(\bm{k}),
\nonumber\\
\text{Re}F_{1}(\bm{k}')&=\text{Re}F_{1}(\bm{k}),
\nonumber\\
\text{Im}F_{1}(\bm{k}')&=-\text{Im}F_{1}(\bm{k}),
\nonumber\\
F^{(1,2)}_{\text{SO}}(\bm{k}')&=-F^{(1,2)}_{\text{SO}}(\bm{k}),
\end{align}
thus
\begin{align}
\widetilde{C}_{6,3}(k_{z})=\tau_{x}\exp\left(i\frac{\pi}{6}\sigma_{z}\right)\exp\left(-i\frac{ck_{z}}{2}\right),
\end{align}
where $\widetilde{C}_{6,3}(k_{z})$ means
the representation of $\widetilde{C}_{6,3}$ in a Bloch basis,
in which the momentum dependent phase factor results from the partial lattice translation along
the $z$ direction.
Finally, for $k_{y}\rightarrow-k_{y}$, we find
\begin{align}
F_{0}(k_{x},-k_{y},k_{z})&=F_{0}(k_{x},k_{y},k_{z}),
\nonumber\\
\text{Re}F_{1}(k_{x},-k_{y},k_{z})&=\text{Re}F_{1}(k_{x},k_{y},k_{z}),
\nonumber\\
\text{Im}F_{1}(k_{x},-k_{y},k_{z})&=-\text{Im}F_{1}(k_{x},k_{y},k_{z}),
\nonumber\\
F^{(1)}_{\text{SO}}(k_{x},-k_{y},k_{z})&=F^{(1)}_{\text{SO}}(k_{x},k_{y},k_{z}),
\nonumber\\
F^{(2)}_{\text{SO}}(k_{x},-k_{y},k_{z})&=-F^{(2)}_{\text{SO}}(k_{x},k_{y},k_{z}),
\end{align}
thus
\begin{align}
\widetilde{M}_{y}(k_{z})=i\tau_{x}\sigma_{y}\exp\left(-i\frac{ck_{z}}{2}\right).
\end{align}
Let us note that $F^{(2)}_{\text{SO}}$ term breaks the glide mirror $\widetilde{M}_{y}$.

On the $k_{z}$ axis with $k_{x}=k_{y}=0$, the Hamiltonian becomes
\begin{align}
H(k_{z})=F_{0}+F_{1}(k_{z})\tau_{x},
\end{align}
from which we find two degenerate eigenstates,
\begin{displaymath}
|\psi_{+1}\rangle=
\frac{1}{\sqrt{2}}\left( \begin{array}{c}
1 \\ 0\\ 1\\0
\end{array} \right),\quad
|\psi_{+2}\rangle=
\frac{1}{\sqrt{2}}\left( \begin{array}{c}
0\\ 1\\0 \\ 1
\end{array} \right),
\end{displaymath}
with the eigenvalue $E_{+}(k_{z})=F_{0}+F_{1}(k_{z})$,
and the other two degenerate eigenstates,
\begin{displaymath}
|\psi_{-1}\rangle=
\frac{1}{\sqrt{2}}\left( \begin{array}{c}
1 \\ 0\\ -1\\0
\end{array} \right),\quad
|\psi_{-2}\rangle=
\frac{1}{\sqrt{2}}\left( \begin{array}{c}
0\\ 1\\0 \\ -1
\end{array} \right),
\end{displaymath}
with the eigenvalue $E_{-}(k_{z})=F_{0}-F_{1}(k_{z})$.
Let us note that $|\psi_{+1}\rangle$, $|\psi_{+2}\rangle$, $|\psi_{-1}\rangle$, $|\psi_{-2}\rangle$
are also the eigenstates of $\widetilde{C}_{6,3}$ with the corresponding eigenvalues
$\exp(-i\frac{ck_{z}}{2}+i\frac{\pi}{6})$, $\exp(-i\frac{ck_{z}}{2}+i\frac{11\pi}{6})$,
$\exp(-i\frac{ck_{z}}{2}+i\frac{7\pi}{6})$, $\exp(-i\frac{ck_{z}}{2}+i\frac{5\pi}{6})$,
respectively.
They are exactly the $\widetilde{C}_{6,3}$ eigenstates [See Eq.~(\ref{eqn:C63_basis})],
which can support a single Dirac point at the zone boundary
on the rotation axis, i.e., at the $A$ point with the momentum $\bm{k}=(0,0,\frac{\pi}{c})$.
The low-energy Hamiltonian near the $A$ point is given by
\begin{align}
H_{A}(\bm{q})&\approx 3t_{2}q_{z}\left(1-\frac{q_{x}^{2}+q_{y}^{2}}{12}\right)\tau_{x}+\frac{t_{2}}{24\sqrt{3}}(3q_{x}^{2}q_{y}-q_{y}^{3})q_{z}\tau_{y}
\nonumber\\
+&\Big[\frac{1}{4}\lambda^{(1)}_{\text{SO}}(3q_{x}q_{y}^{2}-q_{x}^{3})+\frac{3\sqrt{3}}{4}\lambda^{(2)}_{\text{SO}}(3q^{2}_{x}q_{y}-q_{y}^{3})\Big]\tau_{z}\sigma_{z},
\end{align}
where the momentum $\bm{q}$ are measured relative to the $A$ point assuming $a=c=1$
and the constant term $F_{0}$ is dropped.
It is interesting to note that the dispersion on the $(q_{x},q_{y})$ plane
is cubic whereas it is linear along the $q_{z}$ direction.
The cubic dispersion arises due to the angular momentum difference between
the conduction and valence bands,
as indicated in Eq.~(\ref{eqn:C63_basis}).
Thus we obtain a {\it cubic} Dirac point at the $A$ point, which
is protected by the 6-fold screw rotation $\widetilde{C}_{6,3}$.

Finally, let us briefly explain the physical origin of the Dirac point
at the $L$ point with the momentum $\bm{k}=(0,\frac{2\pi}{\sqrt{3}a},\frac{\pi}{c})$.
At the $L$ point, the system is invariant under a set of point group symmetry operations,
which is so-called the little co-group at $L$, $\overline{G}^{L}$.
The little co-group $\overline{G}^{L}$ is generated by three symmetry operations,
a glide mirror $\widetilde{M}_{y}$,
the inversion $P$ and the two-fold rotation about the $y$ axis $C_{2y}$.
Due to the partial lattice translation $\bm{t}=\frac{c}{2}\hat{z}$ involved in $\widetilde{M}_{y}$,
$\widetilde{M}_{y}$ and $P$ do not commute. Namely, we find that
\begin{align}
P\widetilde{M}_{y}:(x,y,z)\rightarrow \left(-x,y,-z-\frac{c}{2}\right),
\end{align}
thus
\begin{align}
P\widetilde{M}_{y}=\left\{PM_{y}|-\frac{c}{2}\hat{z}\right\}.
\end{align}
On the other hand,
\begin{align}
\widetilde{M}_{y}P:(x,y,z)\rightarrow \left(-x,y,-z+\frac{c}{2}\right),
\end{align}
thus
\begin{align}
\widetilde{M}_{y}P=\left\{PM_{y}|\frac{c}{2}\hat{z}\right\}.
\end{align}
Therefore $\widetilde{M}_{y}$ and $P$ anti-commute
at the zone boundary with $k_{z}=\pi/c$,
which guarantees the four-fold degeneracy.
Since two-fold rotation symmetry cannot support a stable Dirac point,
$C_{2y}$ symmetry cannot play an important role here.
Moreover, the $F^{(2)}_{\text{SO}}$ term breaking the $\widetilde{M}_{y}$
vanishes at the $L$ point.
To establish a general theory about the protection of a Dirac point by a glide mirror symmetry
and its associated topological charge is an interesting
research topic, which we leave for future study.

\subsection{\label{sec:diamond} Example 2: a class II Dirac semimetal on a diamond lattice}
\begin{figure}[t]
\centering
\includegraphics[width=8.5 cm]{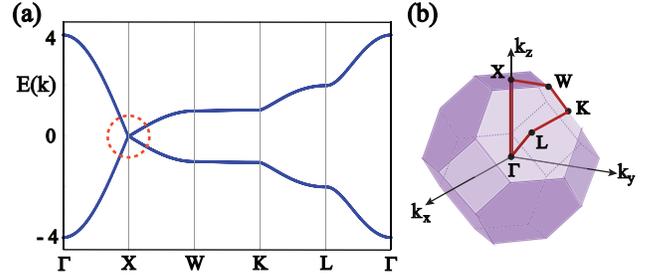}
\caption{
(a) Band structure of the Fu-Kane-Mele model
on a diamond lattice.
There is a Dirac point at the $X$ point.
(b) First Brillouin zone of a diamond lattice.
}
\label{fig:FKM}
\end{figure}
As a second example of class II Dirac semimetal,
let us consider the Fu-Kane-Mele Hamiltonian
on a diamond lattice~\cite{FKM},
\begin{eqnarray}
H=t\sum_{\langle ij \rangle}c_{i}^{\dag}c_{j}
+8i\frac{\lambda_{\text{SO}}}{a^{2}}\sum_{\langle\langle ij \rangle\rangle}c_{i}^{\dag}\bm{\sigma}\cdot(\bm{d}^{1}_{ij}\times \bm{d}^{2}_{ij})c_{j},
\end{eqnarray}
where the first term indicates the nearest-neighbor hopping
and the second term connects the second-nearest-neighbors
with a spin dependent amplitude. $d_{ij}^{1,2}$ are
the two nearest-neighbor bond vectors traversed between
sites $i$ and $j$, and $\sigma_{x,y,z}$
are Pauli matrices indicating the spin degrees of freedom.
$a$ denotes the cubic lattice constant.
In the momentum space, the Hamiltonian becomes
\begin{displaymath}
H(\textbf{k})=
\left( \begin{array}{cc}
\sum_{i=1}^{3}F_{i}\sigma_{i} & F_{0}^{*}\textbf{1}_{2} \\
F_{0}\textbf{1}_{2} & -\sum_{i=1}^{3}F_{i}\sigma_{i}
\end{array} \right),
\end{displaymath}
where $\textbf{1}_{2}$ indicates a $2\times 2$ identity matrix
and $F_{0,1,2,3}$ are given by
\begin{eqnarray}
F_{0}&=&t\big[e^{\frac{ia}{4}(k_{x}+k_{y}+k_{z})}+e^{\frac{ia}{4}(k_{x}-k_{y}-k_{z})}
\nonumber\\
&&+e^{\frac{ia}{4}(-k_{x}+k_{y}-k_{z})}+e^{\frac{ia}{4}(-k_{x}-k_{y}+k_{z})}\big],
\nonumber\\
F_{1}&=&4\lambda_{\text{SO}}\sin\Big(\frac{ak_{x}}{2}\Big)\Big[\cos\big(\frac{ak_{y}}{2}\big)-\cos\big(\frac{ak_{z}}{2}\big)\Big],
\nonumber\\
F_{2}&=&4\lambda_{\text{SO}}\sin\Big(\frac{ak_{y}}{2}\Big)\Big[\cos\big(\frac{ak_{z}}{2}\big)-\cos\big(\frac{ak_{x}}{2}\big)\Big],
\nonumber\\
F_{3}&=&4\lambda_{\text{SO}}\sin\Big(\frac{ak_{z}}{2}\Big)\Big[\cos\big(\frac{ak_{x}}{2}\big)-\cos\big(\frac{ak_{y}}{2}\big)\Big].
\end{eqnarray}
This Hamiltonian exhibits 3D bulk Dirac points at three inequivalent $X$ points $X^{r}=2\pi\hat{r}/a$
where $r=x,y,z$.
Each Dirac point at $X^{r}$  is protected by the 4-fold screw rotation about
$\hat{r}$ axis.

\begin{figure}[t]
\centering
\includegraphics[width=8.5 cm]{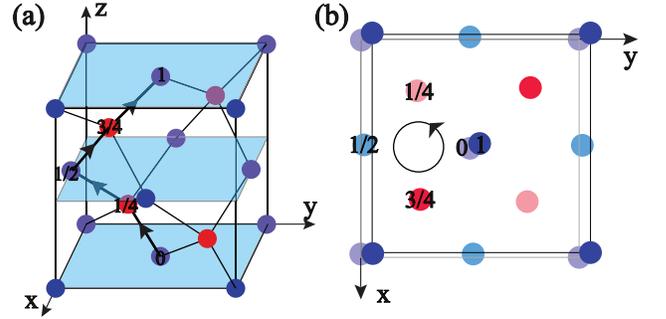}
\caption{
Structure of a diamond lattice.
Two sublattice sites are marked by using different colors.
The arrows indicate a 4-fold screw rotation about the $z$ axis.
}
\label{fig:screw}
\end{figure}
To understand the role of the screw rotation,
let us describe the symmetry of the system.
The space group of the diamond lattice is $Fd3m (O_{h}^{7})$\blue{,}
which contains the 24 symmorphic elements of tetrahedral point group $\bar{4}3m (T_{d})$
and 24 non-symmorphic elements.
The non-symmorphic elements are obtained by compounding
each symmorphic symmetry operation of $T_{d}$ with a translation
along $\bm{t}_{d}=(\frac{a}{4},\frac{a}{4},\frac{a}{4})$, which takes
one sublattice site to another inequivalent sublattice site.
On the other hand, the symmorphic symmetry operation connects sites belonging
to the same sublattice.

At a generic point $\Delta=(0,0,k_{z})$ on the $k_{z}$ axis,
the system has $C_{4v}$ symmetry that is
composed of 5 different symmetry classes with elements $\{E|0\}$,
$\{C_{4}^{2}|0\}$, $2\{C_{4}|\bm{t}_{d}\}$, $2\{iC_{4}^{2}|\bm{t}_{d}\}$, 2$\{iC_{2}'|0\}$,
respectively. It is worth to note that
some symmetry elements contain a partial lattice translation
$\bm{t}_{d}=(\frac{a}{4},\frac{a}{4},\frac{a}{4})$ which
is the characteristic property of the non-symmorphic nature of the diamond lattice space group.
This contrasts with the case of conventional symmorphic cubic lattices such as simple cubic (sc),
face centered cubic (fcc), body centered cubic (bcc) lattices where
the system on the $k_{z}$ axis has
the ordinary $C_{4v}$ group containing only symmorphic point group operations~\cite{grouptheorybook}.

By considering the symmetry of the Hamiltonian, one can easily find the matrix representation
of the screw rotation $\widetilde{C}_{4,1}\equiv\{C_{4z}|\bm{t}_{d}\}=\exp(-i\bm{k}\cdot\bm{t}_{d})\tau_{x}\exp(i\frac{\pi}{4}\sigma_{z})$
where $\exp(-i\bm{k}\cdot\bm{t}_{d})$ represents the translation $\bm{t}_{d}$ of the Bloch state with the momentum $\textbf{k}$,
$\tau_{x}$ indicates the sublattice change due to partial translation, and $\exp(i\frac{\pi}{4}\sigma_{z})$
represents the 4-fold rotation.
Then it is straightforward to confirm that
\begin{eqnarray}
\widetilde{C}_{4,1}H(k_{x},k_{y},k_{z})\widetilde{C}^{-1}_{4,1}=H(-k_{y},k_{x},k_{z}).
\end{eqnarray}

Since $F_{1,2,3}=\text{Im}F_{0}=0$ on the $k_{z}$ axis,
the Hamiltonian on the $k_{z}$ axis becomes
\begin{eqnarray}
H(k_{z})=4t\cos(\frac{ak_{z}}{4})\tau_{x},
\end{eqnarray}
from which we obtain one eigenvalue $E_{+}(\bm{k})=4t\cos(\frac{1}{4}ak_{z})$
with the corresponding eigenvectors
\begin{displaymath}
|\psi_{+1}\rangle=
\frac{1}{\sqrt{2}}\left( \begin{array}{c}
1 \\ 0\\ 1\\0
\end{array} \right),\quad
|\psi_{+2}\rangle=
\frac{1}{\sqrt{2}}\left( \begin{array}{c}
0\\ 1\\0 \\ 1
\end{array} \right),
\end{displaymath}
and the other eigenvalue $E_{-}(\bm{k})=-4t\cos(\frac{1}{4}ak_{z})$
with the eigenvectors
\begin{displaymath}
|\psi_{-1}\rangle=
\frac{1}{\sqrt{2}}\left( \begin{array}{c}
1 \\ 0\\ -1\\0
\end{array} \right),\quad
|\psi_{-2}\rangle=
\frac{1}{\sqrt{2}}\left( \begin{array}{c}
0\\ 1\\0 \\ -1
\end{array} \right).
\end{displaymath}
Let us note that $|\psi_{+1}\rangle$, $|\psi_{+2}\rangle$, $|\psi_{-1}\rangle$, $|\psi_{-2}\rangle$
are also the eigenstates of $\widetilde{C}_{4,1}$ with the eigenvalues
$\exp(-i\frac{ak_{z}}{4}+i\frac{\pi}{4})$, $\exp(-i\frac{ak_{z}}{4}-i\frac{\pi}{4})$,
$\exp(-i\frac{ak_{z}}{4}+i\frac{5\pi}{4})$, $\exp(-i\frac{ak_{z}}{4}-i\frac{5\pi}{4})$,
respectively.
Since the system is $\frac{4\pi}{a}$ periodic along the $k_{z}$ direction,
a single Dirac point protected by $\widetilde{C}_{4,1}$ can be created at the Brillouin zone boundary.
It is straightforward to see that these four eigenvalues
have the same form as
Eq.~(\ref{eqn:screw41_basis}), hence
satisfy the condition for the band crossing
at the Brillouin zone boundary with $k_{z}=\pm2\pi/a$.
Also the band structure of the system along the $k_{z}$ axis
is consistent with Fig.~\ref{fig:screw41}.

Before we close this subsection, let us briefly perform
a group theoretical analysis at the $X$ point.
The little co-group $\overline{G}^{X}$ at the $X$ point with the momentum $\bm{k}=(0,0,\frac{2\pi}{a})$
is generated by three symmetry operators, i.e.,
the four-fold screw rotation $\widetilde{C}_{4,1}$ about the $z$ axis,
the inversion $P$, the two-fold rotation about the $x$ axis $C_{2x}$~\cite{Sigrist,grouptheorybook2}.
One interesting property of the diamond lattice is that
the inversion symmetry $P$ also accompanies a partial translation $\bm{t}_{d}$.
Thus, it is more suitable to use the notation $\widetilde{P}=\{P|\bm{t}_{d}\}$ to indicate
the partial translation associated with the inversion.
Here we describe two interesting physical consequences resulting from the non-symmorphic
nature of $\widetilde{P}$.

Firstly, the partial translation involved in $\widetilde{P}$ does
not affect the commutation relation between $\widetilde{P}$ and
a screw rotation $\widetilde{C}_{N,q}$ on the rotation axis.
This fact can be easily understood by considering the coordinate tranformation
under the combination of $\widetilde{C}_{N,q}=\{C_{N}|\bm{\tau}_{q}\}$ and $\widetilde{P}=\{P|\bm{t}_{P}\}$.
Assuming the $z$ axis is the screw rotation axis, $\widetilde{P}\widetilde{C}_{N,q}$ transforms the coordinate $(x,y,z)$ to
\begin{align}
(-x'-\tau_{q,x}+t_{P,x},-y'-\tau_{q,y}+t_{P,y},-z-\tau_{q,z}+t_{P,z}),
\end{align}
where $x'$ and $y'$ are rotated coordinates
satisfying $x'+iy'=(x+iy)\exp(i2\pi/N)$. Thus we obtain
\begin{align}
\widetilde{P}\widetilde{C}_{N,q}=\{PC_{N}|-\bm{\tau}_{q}+\bm{t}_{P}\}.
\end{align}
On the other hand,
$\widetilde{C}_{N,q}\widetilde{P}$ transforms $(x,y,z)$ to
\begin{align}
(-x'+\tau_{q,x}+t_{P,x}',-y'+\tau_{q,y}+t_{P,y}',-z+\tau_{q,z}+t_{P,z}),
\end{align}
where $t_{P,x}'+it_{P,y}'=(t_{P,x}+it_{P,y})\exp(i2\pi/N)$. Thus we obtain
\begin{align}
\widetilde{C}_{N,q}\widetilde{P}=\{PC_{N}|\bm{\tau}_{q}+\bm{t}_{P}'\}.
\end{align}
Now let us consider a Bloch state $\ket{k_{z}}$ on the rotation axis with the momentum $\bm{k}=(0,0,k_{z})$.
We find
\begin{align}
\widetilde{P}\widetilde{C}_{N,q}\ket{k_{z}}&=\exp\left[ik_{z}(-\tau_{q,z}+t_{P,z})\right]PC_{N}\ket{k_{z}},
\nonumber\\
\widetilde{C}_{N,q}\widetilde{P}\ket{k_{z}}&=\exp\left[ik_{z}(\tau_{q,z}+t_{P,z})\right]PC_{N}\ket{k_{z}},
\end{align}
which shows that the partial translation $t_{P,z}$ associated with the inversion
only provides an overall phase factor, and does not affect the commutation relation
whereas $\tau_{q,z}$ does.
Therefore our theory can also be applied to systems with the inversion $\widetilde{P}$
accompanying a partial translation, as long as
the Dirac point is located on the rotation axis.

However, when the Dirac point is located away from the rotation axis,
the translation $\bm{t}_{P}$ can cause nontrivial physical consequence as well.
For instance, in the $k_{z}=0$ plane, perpendicular to the rotation axis,
a Bloch state $\ket{k_{x},k_{y}}$ satisfies,
\begin{align}
&\widetilde{P}\widetilde{C}_{N,q}\ket{k_{x},k_{y}}
\nonumber\\
&=\exp\big[-ik'_{x}(\tau_{q,x}-t_{P,x})-ik'_{y}(\tau_{q,y}-t_{P,y})\big]PC_{N}\ket{k_{x},k_{y}},
\end{align}
and
\begin{align}
&\widetilde{C}_{N,q}\widetilde{P}\ket{k_{x},k_{y}}
\nonumber\\
&=\exp\big[ik'_{x}(\tau_{q,x}+t_{P,x}')+ik'_{y}(\tau_{q,y}+t_{P,y}')\big]PC_{N}\ket{k_{x},k_{y}}.
\end{align}
where $k'_{x}+ik'_{y}=(k_{x}+ik_{y})\exp(i2\pi/N)$
and we have used the relation $k'_{x}t'_{P,x}+k'_{y}t'_{P,y}=k_{x}t_{P,x}+k_{y}t_{P,y}$.
The point is that since $(t_{P,x},t_{P,y})\neq(t_{P,x}',t_{P,y}')$ due to the rotation,
the partial translation $\bm{t}_{P}$ associated with the inversion
can also modify the commutation relation between the inversion and the rotation symmetries.
Because of this, even a symmorphic rotation symmetry, which is not accompanied by a translation,
can create a Dirac point away from the rotation axis when it is combined with the non-symmorphic inversion symmetry.

For illustration, let us consider
the commutation relation between $C_{2z}$ and $\widetilde{P}=\{P|\bm{t}_{P}\}$
where $\bm{t}_{P}=\bm{t}_{d}=(\frac{a}{4},\frac{a}{4},\frac{a}{4})$.
The Bloch state $\ket{k_{x},k_{y}}$ satisfies,
\begin{align}
&\widetilde{P}C_{2z}\ket{k_{x},k_{y}}
\nonumber\\
&=\exp\big[ik_{x}(-t_{P,x})+ik_{y}(-t_{P,y})\big]PC_{2z}\ket{k_{x},k_{y}},
\end{align}
and
\begin{align}
&C_{2z}\widetilde{P}\ket{k_{x},k_{y}}
\nonumber\\
&=\exp\big[ik_{x}(t_{P,x})+ik_{y}(t_{P,y})\big]PC_{2z}\ket{k_{x},k_{y}}.
\end{align}
Hence at the high symmetry momentum such as $\bm{k}_{1}=(\frac{2\pi}{a},0,0)$
or $\bm{k}_{2}=(0,\frac{2\pi}{a},0)$ where
$\exp\big(ik_{x}t_{P,x}+ik_{y}t_{P,y}\big)=i$,
we obtain $\{\widetilde{P},C_{2z}\}=0$.
This anti-commutation relation can create
a stable Dirac point at $\bm{k}_{1}$ and $\bm{k}_{2}$,
which is again confirmed by $K$-theory analysis in the Appendix \ref{sec:appendix2_3}.
Therefore, although the symmorphic $C_{2}$ symmetry cannot
support a stable Dirac point on the rotation axis,
it can create a stable Dirac point at the zone boundary in the plane perpendicular
to the rotation axis, when the $C_{2}$ is combined with
a non-symmorphic inversion symmetry $\widetilde{P}$.

In fact, as noted before,
there are three symmetry generators, $\widetilde{P}$, $\widetilde{C}_{4,1}$, $C_{2x}$, at the $X$ point with $\bm{k}=(0,0,\frac{2\pi}{a})$
in the diamond lattice.
According to the previous discussion, the Dirac point at $X$ can be protected not only by $\widetilde{P}$ and $\widetilde{C}_{4,1}$ satisfying
$\{\widetilde{P}, \widetilde{C}_{4,1}\}=0$ but also by
$\widetilde{P}$ and $C_{2x}$, which are also anti-commuting $\{\widetilde{P}, C_{2x}\}=0$.
Because of the high crystalline symmetry, the Dirac point
in the diamond lattice is protected by multiple pairs of symmetry operators~\cite{Sigrist}.

\section{\label{sec:discussion} Discussion}

To sum up,
we have studied the topological charge of 3D Dirac semimetals
protected by the time-reversal, the inversion, and the rotation symmetries.
Consideration of topological charges naturally leads to
two different classes of Dirac semimetals, which is consistent with
the previous observation
based on the symmetry constrained minimal Hamiltonian analysis~\cite{BJYang}.
Class I Dirac semimetals are protected by an ordinary
symmorphic rotation symmetry which commutes with the inversion.
Since each eigenstate carries a quantized rotation eigenvalue
on the rotation axis, Dirac points should form a pair having
the opposite topological charges when the system is periodic along
the rotation axis. On the other hand, class II Dirac semimetals
are associated with non-symmorphic screw rotation symmetries.
The eigenvalue of a screw rotation is not quantized on the rotation axis due to
the phase factor induced by a partial lattice translation,
which enables to create a single isolated Dirac point
at the Brillouin zone boundary.

The nonzero topological charge of a Dirac point not only guarantees the stability of the gap-closing point,
but can trigger new types of Lifshitz transitions.
For instance, when two Dirac points merge at the same momentum,
the topological charge of the merging point is given by the summation of their topological charges.
Since the energy dispersion around the gap-closing point strongly depends on its topological charge,
such a merging transition can generate intriguing nodal quasi-particles with novel physical properties~\cite{egmoon,bjyang_QCP,HongYao,Lai}.
Moreover, the presence of a quantized topological charge can be a source of new topological responses.
For instance, it is well-known that the nonzero monopole charge of Weyl points
induces novel topological responses in Weyl semimetals~\cite{Wan}.
Recent theoretical studies of interesting topological responses in Dirac semimetals~\cite{Ramamurthy,Nematic}
may imply nontrivial role of topological charges in these systems.

Up to now, two materials (Na$_{3}$Bi and Cd$_{3}$As$_{2}$)
belonging to the class I are discovered and extensively studied whereas
class II Dirac semimetals are not uncovered yet.
Though there are some hypothetical candidate materials proposed theoretically~\cite{Young1,Young2},
all of them are chemically unstable because the metallic ion in each candidate compound
is required to have a lone-pair valence electron to locate the Fermi level
at the Dirac point~\cite{Cava}. In this respect, synthesizing class II Dirac semimetals
is a challenging problem in material science which should be properly addressed in near future.

We believe that class II Dirac semimetals
are as important as class I Dirac semimetals in the following sense.
In the case of class I Dirac semimetals,
the Dirac points are created by a band inversion, hence
the intrinsic properties of 3D Dirac particles can be observed
only within the narrow energy scale associated with the band inversion~\cite{Cd3As2_Exp3}.
Because of this, if the competing energy scales, such as
the Fermi energy due to doped carriers or the disorder-induced
broadening, are comparable to the band inversion energy,
the intrinsic properties of 3D Dirac particles can be easily washed out.
However, in the case of class II Dirac semimetals,
the energy scale of the Dirac dispersion is simply given
by the bandwidth of the system (roughly in the order of a few eV),
which obviously provides a better playground to study the intrinsic properties
of 3D Dirac particles.

Secondly, we would like to draw attention to class II Dirac semimetals
as potential novel topological states.
The fact that the Fu-Kane-Mele model is a canonical model
to construct a 3D $Z_{2}$ topological insulator implies the intrinsic
topological nature of the associated Dirac semimetal state.
In particular, in the present paper, we have demonstrated
that the presence of a single Dirac point on the rotation axis is
unnatural in consideration of the Nielsen-Ninomiya theorem, and, in fact,
the projective nature of the screw rotation symmetry plays an essential role
to circumvent the doubling of Dirac points.
Although the discrete nature of the rotation symmetry
should be distinct from the continuous U(1) symmetry associated
with the original Nielsen-Ninomiya theorem,
the mechanism leading to circumventing the fermion number doubling
shares the common origin, i.e., assigning a non-quantized quantum number to fermions.
To reveal the topological properties of class II Dirac semimetals would definitely be an exciting
theoretical problem which we leave for future studies.

Finally, we would like to point out
that there are a class of Dirac semimetals
which are not completely treated in our classification scheme.
Let us note that, in both class I and II Dirac semimetals considered in the present work,
Dirac points are located on the rotation axis.
However, rotation symmetries can also protect
a Dirac point which is away from the rotation axis.
For instance, we have shown in Sec.~\ref{sec:diamond} that
the symmorphic $C_{2}$ rotation can protect a Dirac point
which is not on the rotation axis,
when it is combined with a non-symmorphic inversion symmetry.
Moreover, the tight-binding model on a hcp lattice considered in
Sec.~\ref{sec:hcp} indicates that glide mirror symmetries can also
give rise to symmetry-protected Dirac points.
To find a systematic way to classify these different types of Dirac semimetals
would also be an important problem for future research.

\acknowledgments
BJY is supported from the Japan Society for the Promotion of Science (JSPS)
through the `Funding Program for World-Leading Innovative R\&D on Science and Technology (FIRST Program),
and Grant-in-Aids for Scientific Research (Kiban (S), No. 24224009) from the
Ministry of Education, Culture, Sports, Science and Technology (MEXT).
AF is grateful for support by Grants-in-Aid from
the Japan Society for Promotion of Science (Grant No.15K05141)
and by the RIKEN iTHES Project.

\appendix

\section{Absence of a stable Dirac point in systems with time-reversal and inversion symmetries only}
Here we prove that
stable Dirac points do not exist in systems having only the time-reversal ($T$) and inversion ($P$) symmetries.
For this, we distinguish two cases: one is when
the Dirac point is located at a generic momentum point, and the other is when
the Dirac point is located at a time-reversal invariant momentum (TRIM).
This distinction is necessary because the
symmetry associated with the Dirac point differs depending on the position of the Dirac point
in the momentum space.
In each case, the stability of a Dirac point is determined by using $K$-theory~\cite{Kitaev, Morimoto_general}.

\subsection{A Dirac point located at a generic momentum}
When the Dirac point locates at a generic momentum,
the combination of $T$ and $P$ is the only symmetry satisfied around the Dirac point.
In general, a $PT$ symmetric system satisfies
\begin{eqnarray}\label{eqn:PTinvariance}
(PT)H(\bm{k})(PT)^{-1}=H(\bm{k}),
\end{eqnarray}
where the anti-unitary $PT$ symmetry satisfies $(PT)^{2}=-1$, which is coming from
\begin{eqnarray}
P^{2}=1,\quad T^{2}=-1,\quad [P,T]=0,
\end{eqnarray}
in electronic systems.

The stability of the Dirac point can be understood by using K theory approach.
We consider a Dirac point locating at a generic momentum $\bm{k}_{0}=(k_{x}^{0},k_{y}^{0},k_{z}^{0})$.
The effective Hamiltonian describing
the low energy excitation around the Dirac point is given by
\begin{eqnarray}
H_{D}=(k_{x}-k_{x}^{0})\gamma_{x}+(k_{y}-k_{y}^{0})\gamma_{y}+(k_{z}-k_{z}^{0})\gamma_{z}+m\gamma_{0},\nonumber
\end{eqnarray}
where $\gamma_{0,x,y,z}$ are gamma matrices satisfying the anticommutation relations $\{\gamma_{i},\gamma_{j}\}=2\delta_{i,j}$,
and $m$ indicates a possible symmetry-preserving Dirac mass term.
The presence (absence) of the symmetry-preserving Dirac mass term $m$ indicates
the instability (stability) of the Dirac point.
From Eq.~(\ref{eqn:PTinvariance}), we obtain
\begin{eqnarray}
[\gamma_{x},PT]=[\gamma_{y},PT]=[\gamma_{z},PT]=[\gamma_{0},PT]=0.
\end{eqnarray}

To confirm the existence or absence
of the Dirac mass term $m$,
let us define a real Clifford algebra generated by $PT$ and $\gamma_{0,x,y,z}$, which is given by
\begin{eqnarray}\label{eqn:algebra1}
Cl_{0,4}\otimes Cl_{2,0}=\{;\gamma_{x},\gamma_{y},\gamma_{z},\gamma_{0}\}\otimes\{PT,JPT;\},
\end{eqnarray}
where $Cl_{p,q}$ indicates a real Clifford algebra with
$p+q$ generators $\{e_{1},e_{2},...,e_{p};e_{p+1},e_{p+2},...,e_{p+q}\}$ satisfying
\begin{align}
\{e_{i},e_{j}\}&=0,\quad i\neq j
\nonumber\\
e_{i}^{2}&=
\begin{cases}
-1, \quad 1\leq i \leq p,
\nonumber\\
+1, \quad p+1\leq i \leq p+q.
\end{cases}
\end{align}
The algebra in Eq.~(\ref{eqn:algebra1}) can be easily obtained by considering the following relations:
\begin{eqnarray}
&(i)&[\gamma_{x},PT]=[\gamma_{y},PT]=[\gamma_{z},PT]=[\gamma_{0},PT]=0,
\nonumber\\
&(ii)&[\gamma_{x},J]=[\gamma_{y},J]=[\gamma_{z},J]=[\gamma_{0},J]=0,
\nonumber\\
&(iii)&\{\gamma_{i},\gamma_{j}\}=2\delta_{i,j},
\nonumber\\
&(iv)&\{PT,JPT\}=0,~(PT)^{2}=(JPT)^{2}=-1,
\end{eqnarray}
where the symbol $J$ indicating the pure imaginary number $i$
is introduced to construct a real Clifford algebra.

The existence or absence of the Dirac mass $m\gamma_{0}$
can be judged by considering the following extension problem:
\begin{eqnarray}
&&\{;\gamma_{x},\gamma_{y}\}\otimes\{PT,JPT;\}
\nonumber\\
&&\rightarrow
\{;\gamma_{x},\gamma_{y},\gamma_{z}\}\otimes\{PT,JPT;\},
\end{eqnarray}
i.e.,
\begin{eqnarray}
Cl_{0,2}\otimes Cl_{2,0}\rightarrow
Cl_{0,3}\otimes Cl_{2,0}.
\end{eqnarray}
Namely, the topological classification of $\gamma_{z}$ determines
the topological nature of the Dirac point. This is because
the topologically trivial classification of $\gamma_{z}$ implies
the existence of another gamma matrix such as $\gamma_{0}$
which anticommutes with the three generators $\gamma_{x,y,z}$ (i.e., a mass term exists)
whereas the topologically nontrivial classification of $\gamma_{z}$
implies the absence of $\gamma_{0}$, thus the topologically nontrivial
nature of the Dirac point.
Generally, in the classification scheme with Clifford algebera,
the existence condition of a particular generator $e_{i}$ (Dirac mass term)
is equivalent to the classification of another generator of the same type as $e_{i}$
in the Clifford algebra in which $e_{i}$ is removed.~\cite{Morimoto_Z2}

The extension problem
\begin{eqnarray}
Cl_{0,2}\otimes Cl_{2,0}\rightarrow
Cl_{0,3}\otimes Cl_{2,0}
\end{eqnarray}
is equivalent to
\begin{eqnarray}
Cl_{4,0}\rightarrow
Cl_{5,0}
\end{eqnarray}
due to the relation
\begin{eqnarray}
Cl_{p,q}\otimes Cl_{2,0} \simeq Cl_{q+2,p}.
\end{eqnarray}
Since the classifying space for the extension
$Cl_{p,q}\rightarrow Cl_{p+1,q}$ is given by $R_{p+2-q}$,
the classifying space for the extension $Cl_{4,0}\rightarrow Cl_{5,0}$
is $R_{6}=Sp(n)/U(n)$ with a sufficiently large integer $n$.
Since $\pi_{0}(R_{6})=0$, the space of possible representation for $\gamma_{z}$
is singly connected. Namely, a Dirac mass term always exists, hence
the Dirac point is unstable.

\subsection{A Dirac point locating at a TRIM}
On the other hand, when the Dirac point locates at a TRIM,
both $P$ and $T$ are the symmetry of the Dirac point.
To understand the stability of the Dirac point, we consider the following Dirac Hamiltonian
\begin{eqnarray}
H_{D}=k_{x}\gamma_{x}+k_{y}\gamma_{y}+k_{z}\gamma_{z}+m\gamma_{0},
\end{eqnarray}
where the matrices $\gamma_{0,x,y,z}$ satisfy $\{\gamma_{i},\gamma_{j}\}=2\delta_{i,j}$.
$T$ and $P$ symmetries require
\begin{eqnarray}
\{\gamma_{x},T\}&=&\{\gamma_{y},T\}=\{\gamma_{z},T\}=[\gamma_{0},T]=0,
\nonumber\\
\{\gamma_{x},P\}&=&\{\gamma_{y},P\}=\{\gamma_{z},P\}=[\gamma_{0},P]=0,
\end{eqnarray}
where
\begin{eqnarray}
P^{2}=1, T^{2}=-1, [T,P]=0.
\end{eqnarray}
The Clifford algebra generated by $\gamma_{0,x,y,z}$, $T$, $P$, $J$
is given by
\begin{eqnarray}
\{T,JT,J\gamma_{0};\gamma_{x},\gamma_{y},\gamma_{z},P\gamma_{x}\gamma_{y}\gamma_{z}\},
\end{eqnarray}
The existence or absence of the Dirac mass $m\gamma_{0}$
can be judged by considering the following extension problem:
\begin{eqnarray}
&&\{JT;\gamma_{x},\gamma_{y},\gamma_{z},P\gamma_{x}\gamma_{y}\gamma_{z}\}
\nonumber\\
&&\rightarrow
\{T,JT;\gamma_{x},\gamma_{y},\gamma_{z},P\gamma_{x}\gamma_{y}\gamma_{z}\}
\end{eqnarray}
i.e.,
\begin{eqnarray}
Cl_{1,4}\rightarrow
Cl_{2,4},
\end{eqnarray}
the corresponding classifying space is $R_{p+2-q}=R_{-1}\simeq R_{7}$.
Since $\pi_{0}(R_{7})=0$, a Dirac mass term always exists, hence
the Dirac point is unstable.
Therefore, independent of the location of the Dirac point
in the momentum space, the system with only $T$ and $P$ symmetries
cannot support a stable Dirac point.

\section{The stability of Dirac points in $C_{2}$ symmetric systems}
In $C_{2}$ invariant systems, since all the symmetry operators squared become $\pm 1$,
the classification scheme based on Clifford algebras can be applied.

\subsection{When $[C_{2},P]=0$}
In systems with the symmorphic $C_{2}$ rotation satisfying $(C_{2})^{2}=-1$,
the Dirac point can be located at a generic momentum $\bm{k}_{0}=(0,0,k_{z}^{0})$ on the rotation axis ($z$ axis).
We consider the following massive Dirac Hamiltonian,
\begin{align}
H(k)=k_{x}\gamma_x  + k_{y} \gamma_y  + (k_{z}-k_{z}^{0})\gamma_{z} + m\gamma_0 ,
\end{align}
which satisfy the following relations,
\begin{align}
(PT)H(\bm{k})(PT)^{-1}&=H(\bm{k}),\\
C_{2}H(k_{x},k_{y},k_{z})(C_{2})^{-1}&=H(-k_{x},-k_{y},k_{z}).
\end{align}
From this, we obtain
\begin{align}
[\gamma_{x},PT]=[\gamma_{y},PT]=[\gamma_{z},PT]=[\gamma_{0},PT]=0,\\
\{\gamma_{x},C_{2}\}=\{\gamma_{y},C_{2}\}=[\gamma_{z},C_{2}]=[\gamma_{0},C_{2}]=0.
\end{align}
Then we can determine the Clifford algebra generated by the gamma matrices
in the Dirac Hamiltonian
and the relevant symmetry operators.
The resulting Clifford algebra is
\begin{align}
&Cl_{6,0}\tensor Cl_{0,1}
\nonumber\\
&=\{
PT, JPT, J\gamma_x, J\gamma_y, J\gamma_z, J\gamma_0 ;
\}
\tensor \{ ;\gamma_x \gamma_y C_2 \}.
\end{align}
The relevant extension problem is
\begin{align}
Cl_{4,0}\tensor Cl_{0,1} \to Cl_{5,0}\tensor Cl_{0,1},
\end{align}
for which the classifying space is given by $R_6 \times R_6$.
From its zeroth homotopy group,
we find the topological charge as $\pi_0(R_6 \times R_6)=0$.
Therefore a Dirac point cannot carry a nontrivial topological charge,
which is consistent with the absence of a topological invariant found before.
(See Table I.)

\subsection{When $\{\widetilde{C}_{2},P\}=0$}
Now we consider a two-fold screw rotation $\widetilde{C}_{2}$ satisfying $(\widetilde{C}_{2})^{2}=1$.
Since the Dirac point locates at a TRIM, both the $P$ and the $T$ are
the symmetry of the Dirac point.
The commutation relations relevant to this problem are as follows.
\begin{align}\label{eqn:C2P_1}
\{\gamma_{x},T\}&=\{\gamma_{y},T\}=\{\gamma_{z},T\}=[\gamma_{0},T]=0,
\nonumber\\
\{\gamma_{x},P\}&=\{\gamma_{y},P\}=\{\gamma_{z},P\}=[\gamma_{0},P]=0,
\nonumber\\
\{\gamma_{x},\widetilde{C}_{2}\}&=\{\gamma_{y},\widetilde{C}_{2}\}=[\gamma_{z},\widetilde{C}_{2}]=[\gamma_{0},\widetilde{C}_{2}]=0,
\end{align}
and
\begin{align}\label{eqn:C2P_2}
[T,P]=0, ~[\widetilde{C}_{2},T]=0,~ \{\widetilde{C}_{2},P\}=0,
\end{align}
in which
\begin{align}\label{eqn:C2P_3}
P^{2}=1, ~T^{2}=-1, ~(\widetilde{C}_{2})^{2}=1.
\end{align}
The relevant Clifford algebra of gamma matrices and symmetry operators is given by
\begin{align}
Cl_{3,5}=\{ T, JT, J\gamma_{0}; \gamma_x, \gamma_y, \gamma_z, P\gamma_x\gamma_y\gamma_z, \widetilde{C}_2 P \gamma_z \}.
\end{align}
The existence condition of the Dirac mass term
is determined by the extension problem
\begin{align}
Cl_{1,5} \to Cl_{2,5},
\end{align}
for which the classifying space is $R_6$.
From $\pi_0(R_6)=0$,
we see that the $\widetilde{C}_{2}$ symmetry cannot protect a Dirac point
consistent with Table III.

\subsection{\label{sec:appendix2_3} When $\{C_{2},\widetilde{P}\}=0$}
This is relevant to the case when
the Dirac point is located at a TRIM
in the plane perpendicular to the rotation axis.
The commutation relations relevant to this problem are as follows.
\begin{align}
\{\gamma_{x},T\}&=\{\gamma_{y},T\}=\{\gamma_{z},T\}=[\gamma_{0},T]=0,
\nonumber\\
\{\gamma_{x},\widetilde{P}\}&=\{\gamma_{y},\widetilde{P}\}=\{\gamma_{z},\widetilde{P}\}=[\gamma_{0},\widetilde{P}]=0,
\nonumber\\
\{\gamma_{x},C_{2}\}&=\{\gamma_{y},C_{2}\}=[\gamma_{z},C_{2}]=[\gamma_{0},C_{2}]=0,
\end{align}
and
\begin{align}
[T,\widetilde{P}]=0,~ [C_{2},T]=0,~ \{C_{2},\widetilde{P}\}=0,
\end{align}
in which
\begin{align}
\widetilde{P}^{2}=-1,~ T^{2}=-1,~ (C_{2})^{2}=-1.
\end{align}
Then the relevant Clifford algebra of gamma matrices and symmetry operators is given by
\begin{align}
Cl_{3,5}=\{ T, JT, J\gamma_{0}, \widetilde{P}\gamma_x\gamma_y\gamma_z; \gamma_x, \gamma_y, \gamma_z, C_2 \widetilde{P} \gamma_z \}.
\end{align}
The existence condition of the Dirac mass term
is determined by the extension problem
\begin{align}
Cl_{2,4} \to Cl_{3,4},
\end{align}
for which the classifying space is $R_0$.
From $\pi_0(R_0)=\mathbb{Z}$,
we see that the symmorphic $C_{2}$ symmetry can protect a Dirac point
when it is combined with the non-symmorphic inversion symmetry $\widetilde{P}$.
Here the location of the Dirac point is not on the rotation axis but
at a TRIM on the plane perpendicular to the rotation axis because
the anti-commutation relation $\{C_{2},\widetilde{P}\}=0$ can be satisfied only away from the rotation axis.

\section{The stability of 2D Dirac points in systems with two-fold screw rotations}
Recently, Young and Kane proposed a theory~\cite{2DDirac}
about 2D Dirac points located at a TRIM
on the Brilluoin zone boundary.
One interesting finding in their work
is that a two-fold screw rotation can
protect a Dirac point at a TRIM
on the rotation axis,
which is forbidden in the case of 3D systems.
To confirm their claim, let us check the stability of the Dirac point
by using $K$ theory.

Let us consider a Dirac Hamiltonian at the zone boundary.
\begin{align}
H(\bm{k})=k_{x}\gamma_{x}+k_{y}\gamma_{y}+m\gamma_{0},
\end{align}
where $\gamma_{0,x,y}$ are mutually anti-commuting.
Under $P$, $T$, $\widetilde{C}_{2y}$ symmetry satisfying
\begin{align}
&[P,T]=[\widetilde{C}_{2y},T]=0,~\{\widetilde{C}_{2y},P\}=0,
\nonumber\\
&P^{2}=1,~T^{2}=-1,~\widetilde{C}_{2y}^{2}=1,
\end{align}
the gamma matrices satisfy
\begin{align}
&\{\gamma_{x},T\}=\{\gamma_{y},T\}=[\gamma_{0},T]=0,
\nonumber\\
&\{\gamma_{x},P\}=\{\gamma_{y},P\}=[\gamma_{0},P]=0,
\nonumber\\
&\{\gamma_{x},\widetilde{C}_{2y}\}=[\gamma_{y},\widetilde{C}_{2y}]=[\gamma_{0},\widetilde{C}_{2y}]=0,
\end{align}
The relevant Clifford algebra of gamma matrices and symmetry operators is given by
\begin{align}
Cl_{4,2}\otimes Cl_{1,0}=\{T,TJ,J\gamma_{0},\widetilde{C}_{2y}\gamma_{x};\gamma_{x},\gamma_{y}\}\otimes\{P\gamma_{x}\gamma_{y};\}
\end{align}
The existence of the Dirac mass term $m\gamma_{0}$ is determined
by the extension problem
\begin{align}
Cl_{2,2}\otimes Cl_{1,0}\rightarrow Cl_{3,2}\otimes Cl_{1,0}.
\end{align}
Since the extra generator $P\gamma_{x}\gamma_{y}$ commutes with
all the other generator, and satisfies $(P\gamma_{x}\gamma_{y})^{2}=-1$,
the above extension problem is rearranged in the following way,
\begin{align}
Cl_{4}\rightarrow Cl_{5},
\end{align}
for which the classifying space is $C_{4}\cong C_{0}$.
Namely, the extra generator $P\gamma_{x}\gamma_{y}$ deforms
the original real Clifford algebra extension problem
to a complex Clifford algebra extension problem~\cite{Lu}.
From $\pi_{0}(C_{0})=\mathbb{Z}$,
we see that $\widetilde{C}_{2}$ can protect a 2D Dirac point
with topological charge $Z$ on the rotation axis.

\end{document}